%% file: proxy2.tex
\documentclass[12pt]{article}
\usepackage{amssymb,color}
\usepackage{amsmath}
\usepackage{graphicx}
\usepackage{natbib}
\usepackage{subfigure}
\usepackage{latexsym}
\usepackage{bm}
\usepackage[latin1]{inputenc}
\usepackage{bbm}
\usepackage{mathrsfs,accents}
\usepackage{amsfonts}
\usepackage{amssymb}
\usepackage{rotating}
\usepackage{lscape}
\usepackage{amstext}
\usepackage{subfigure}
\usepackage{makeidx}
\usepackage{multicol}
\usepackage{multirow}

\makeatletter

\renewcommand\section{\@startsection {section}{1}{\z@}%
  {-1.5ex \@plus -1ex \@minus -.2ex}%
  {1.3ex \@plus.2ex}%
  {\normalfont\large\bfseries}}
\renewcommand\subsection{\@startsection{subsection}{2}{\z@}%
  {-1.25ex\@plus -1ex \@minus -.2ex}%
  {0.5ex \@plus .2ex}%
  {\normalfont\normalsize\bfseries}}
\renewcommand\subsubsection{\@startsection{subsubsection}{3}{\z@}%
  {-1.25ex\@plus -1ex \@minus -.2ex}%
  {0.5ex \@plus .2ex}%
 {\normalfont\normalsize\itshape}}

\makeatother

\ifx\pdfoutput\undefined
\else
\fi

\newcommand{\cov}{\mathrm{cov}}
\newcommand{\var}{\mathrm{var}}

\DeclareMathAlphabet{\mathbi}{OML}{cmm}{b}{it}

\newcommand{\cum}{\mathrm{cum}}
\newcommand{\Ex}{\mathrm{E}}

\newtheorem{theorem}{Theorem}[section]
\newtheorem{lemma}{Lemma}[section]
\newtheorem{assumption}{Assumption}[section]
\newtheorem{corollary}{Corollary}[section]
\newtheorem{example}{Example}[section]
\newtheorem{remark}{Remark}[section]
\newcommand{\AScon}{\stackrel{\textrm{a.s.}}{\rightarrow}}
\newcommand{\Pcon}{\stackrel{\mathcal{P}}{\rightarrow}}
\newcommand{\Dcon}{\stackrel{\mathcal{D}}{\rightarrow}}

\begin{document}

\title{{\bf Orthogonal samples for estimators in time series}}
\author{Suhasini Subba Rao\\
Department of Statistics, Texas A\&M University \\
College Station, TX, U.S.A.\\
{\tt suhasini@stat.tamu.edu}} 
\date{\today}

\maketitle

\begin{abstract}
Inference for statistics of a stationary time series often
involve nuisance parameters and sampling distributions that  are
difficult to estimate. In this paper,  we propose the
method of orthogonal samples, which can be used to address some of these
issues. For a broad class of statistics,
an orthogonal sample is constructed 
through a slight modification of the original statistic, such that
it shares similar distributional properties as the centralised 
statistic of interest. We use the orthogonal sample to estimate nuisance
parameters of weighted average periodogram estimators and $L_{2}$-type
spectral statistics. Further, the orthogonal sample is utilized to
estimate the finite sampling distribution of various test statistics
under the null hypothesis. The proposed method is simple and
computationally fast to implement. The viability of the method is
illustrated with various simulations. 

\end{abstract}


\input{1_introduction}


\input{2_estimationv2}

\input{4_testing}

\input{4_a_simulations}

\input{3_selection}

\begin{figure}[h!]
\centering
\includegraphics[width = 12cm, height = 8cm]{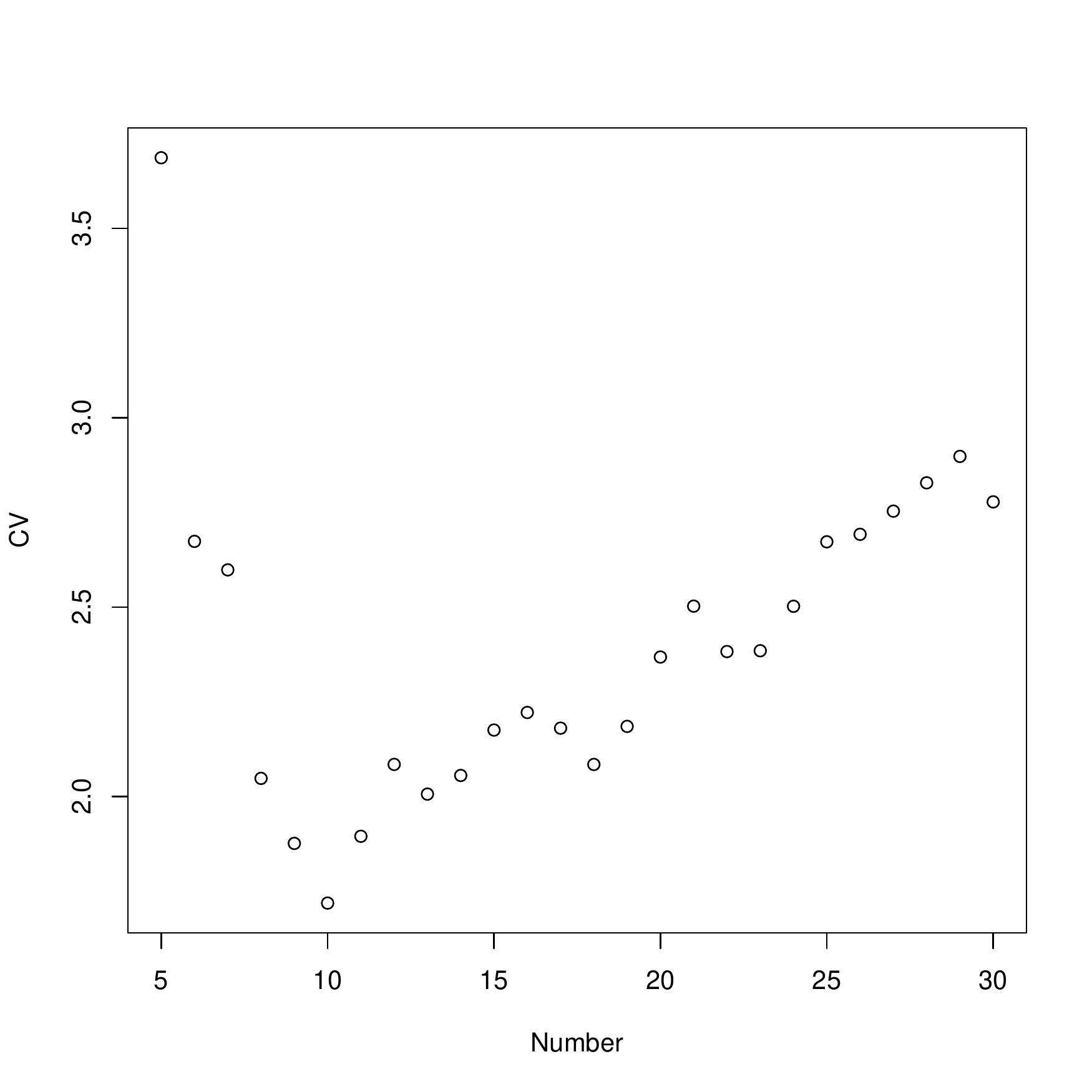}
\caption{The average squared criterion for the sample autocovariance
  function $\mathcal{C}_{e^{i\cdot}}(M)$ ($p=4$) at lag one for the Gaussian autoregressive time
  series $X_{t} = 1.5X_{t-1}-0.75X_{t-2}+\varepsilon_{t}$ where $T=200$.\label{fig:1}}
\end{figure}

\section*{Concluding Remarks}

In this paper we have introduced the method of orthogonal samples for
estimating nuisance parameters in time series. We have applied the
method to some popular statistics in time series. 
Through simulations we have compared this method to well established
methods in the literature, such as
the bootstrap. Our simulations demonstrate that 
orthogonal samples does not consistently
outperform the bootstrap (there were situations where the bootstrap
performed better and others where orthogonal samples performed
better). However, this was not the intention of the paper. 
An advantage of the proposed method is that it is
extremely fast to implement. It can be argued that since computing power is growing
year on year, this is no longer an important issue. However, data sets are 
also increasingly size. Given the growing complexity of modern data
sets, it is the author's view that orthogonal samples may be worth further investigation.



 \subsection*{Acknowledgements}

This work has been partially supported by the National Science
Foundation, DMS-1513647.

\bibliography{bibNSF2014}

\newpage

\appendix

\input{appendix2}

\bibliographystyle{plainnat}

\end{document}

%% file: 1_introduction.tex
\section{Introduction}

In classical statistics, given the correct distribution it is often
possible to define estimators and pivotal quantities which do not
depend on any nuisance parameters, examples include the studentized
t-statistic and log-likelihood ratio statistic. In time series
analysis, due to dependence in the data and that the underlying
distribution of the process is usually unknown, such statistics rarely arise.
However, for inference it is necessary to estimate the
variance of statistic which can often involve a complicated function
of higher order cumulants. For the past 30 years, the standard approach
to the estimation of nuisance parameters and finite sample distributions of
statistics is to use the bootstrap.  This is a simple method
for mimicing the behaviour of the time series. 
There exists many methods for constructing the
bootstrap. Classical examples include the block-type bootstrap 
(see \cite{p:kun-89},
\cite{p:pol-rom-94}, \cite{p:rom-tho-96}, \cite{b:pol-rom-99},
\cite{b:lah-03}, \cite{p:kir-pol-11} and
\cite{p:kri-lah-12}), fixed-b
bootstrap, which accounts for the influence of bandwidth in the block
bootstrap (see \cite{p:kie-05} and \cite{p:sha-13}), 
sieve bootstrap (see \cite{p:kre-92}, \cite{p:kre-10} and \cite{p:jen-pol-13})
frequency domain bootstrap (\cite{p:hur-87},
\cite{p:fra-har-92} and \cite{p:dah-jan-96})
linear process bootstrap (\cite{p:pol-10} and \cite{p:jen-pol-15})
  and moving average
bootstrap (see \cite{p:pap-16}).  An alternative,
is the method of self-normalisation proposed  in 
\cite{p:lob-01} and \cite{p:sha-09}, where 
 the limiting distribution is non-standard but free of nuisance parameters.

The purpose of this paper is to propose an alternative method to
nuisance parameters estimation and pivotal statistics. 
We make no claims that the proposed method is better
than any of the excellent methods mentioned above, but we believe it
may be worth further investigation. 
In many respects our approach is very
classical. It is motivated by Fisher's definition of an ancillary
variable from the 1930s and by the innovations in spectral analysis
for time series developed during the 1950's and 1960's. 
An ancillary variable, is a statistic
whose sampling distribution does not depend on the parameters of
interest yet holds important information about the statistic of
interest. For example, if $\{X_{i}\}_{i=1}^{n}$ are iid random
variables with mean $\mu$ and variance $\sigma^{2}$
and 
$\bar{X}$ is the sample mean, then $X_{i}-\bar{X}$ can be considered
as an ancillary variable since its sampling distribution does not
change with $\mu$, however, since $\var[X_{i}-\bar{X}] = (n-1)\sigma^{2}/n$
it does contain information about the
variance $\sigma^{2}$. Thus the ancillary variables, $\{X_{i} - \bar{X}\}_{i=1}^{n}$,
are used to estimate the variance of the sample
mean. Ancillary variables rarely occur in time series analysis, however,
our aim is to show that several estimators give rise to
\emph{asymptotic} ancillary variables, which can be used to
estimate the variance of the estimator of interest and construct
pivotal quantities. Since the asymptotic ancillary variables 
constructed in this paper are uncorrelated to each other in this paper we call them an \emph{orthogonal
sample}. 

To illustrate the proposed method we consider a well known example in time series, where 
implicitly the notion of an orthogonal sample. Let 
$\{X_{t}\}$ be a stationary, short memory, time series with mean
$\mu$, autocovariance function
$\{c(j)\}$ and spectral density function $f(\omega) =
\frac{1}{2\pi}\sum_{r\in \mathbb{Z}}c(j)\exp(ij\omega)$ and $i =\sqrt{-1}$. We observe
$\{X_{t}\}_{t=1}^{T}$ and use the sample mean $\bar{X}_{T}$ as the
estimator of the spectral density function. It is well know that the 
variance of the sample mean is asymptotically equal to the long run variance 
$\var[\sqrt{T}\bar{X}_{T}] \approx \sum_{j\in \mathbb{Z}}c(j)$.  We
recall that $\sum_{j \in \mathbb{Z}} c(j) = 2\pi f(0)$, thus
estimation of the long run variance is equivalent to estimating the 
spectral density function at frequency zero. A classical estimator of the
spectral density is the local average of the periodogram as a spectral density
function (see \cite{p:bar-50} and \cite{p:par-57}). Applying this to long run variance
estimation, this means using $2\pi \widehat{f}(0)$ as an estimator of $2\pi f(0)$,
where
\begin{eqnarray*}
\widehat{f}(0) = \frac{2\pi}{M} \sum_{k=1}^{M}|J_{T}(\omega_{k})|^{2}
\textrm{ with }
J_{T}(\omega_{k}) = \frac{1}{\sqrt{2\pi
    T}}\sum_{t=1}^{T}X_{t}\exp(it\omega_{k}) \textrm{ and }\omega_{k}
= \frac{2\pi k}{T}.
\end{eqnarray*}
We now take a step back and consider this estimator from a slightly
different perspective, which fits with the notion of an ancillary variable,
discussed above. We observe (i) $\sqrt{2\pi}J_{T}(0) =
\sqrt{T}\bar{X}$ (ii) $\Ex[J_{T}(\omega_{k})] = 0$ if $1\leq k < T/2$,
$\cov[J_{T}(\omega_{k_1}),J_{T}(\omega_{k_2})] = O(T^{-1})$
(if $k_1\neq k_2$) and (iii) if $\omega_{k}$ is in a "neighbourhood" of zero
$\var[J_{T}(\omega_{k})]\approx f(0)$. Thus $M<<T$
$\{J_{T}(\omega_{k})\}_{k=1}^{M}$ can be considered as an orthogonal
sample to the sample mean $\sqrt{2\pi }J_{T}(0)$; it contains no mean
information but shares the same (asymptotic) variance as the sample
mean. Furthermore, if it can be shown that the random vector 
$\{J_{T}(\omega_{k})\}_{k=0}^{M}$ is asymptotically normal (cf.
\cite{b:bro-dav-91}), then for fixed $M$ we have the asymptotically
pivotal statistic
\begin{eqnarray}
\label{eq:pivotal-mean}
\frac{\sqrt{T}(\bar{X} -
  \mu)}{\sqrt{\frac{1}{M}\sum_{k=1}^{M}|J_{T}(\omega_{k})|^{2}}}
\Dcon t_{2M}, \textrm{ as }T\rightarrow \infty
\end{eqnarray}
where $t_{2M}$ denotes a t-distribution with $2M$ degrees of freedom.

The objective in this paper is to generalize the notion outlined above
to a broad class of estimators. Our main focus is the class of weighted average periodogram 
estimators which take the form
\begin{eqnarray}
\label{eq:ATl}
A_{T}(\phi) =
\frac{1}{T}\sum_{k=1}^{T}\phi(\omega_{k})|J_{T}(\omega_{k})|^{2},
\end{eqnarray}
which was first introduced in \cite{p:par-57} and includes the 
sample autocovariance function, 
spectral density estimators and Whittle 
likelihood estimators.  We briefly summarize  the paper.
In Section \ref{sec:quad} we define the orthogonal sample
associated with $A_{T}(\phi)$ which shares similar properties to the
centralised $A_{T}(\phi)$, in particular the same variance.
Using the orthogonal sample we obtain an estimator of the
variance and define an asymptotically pivotal statistic analogous to 
(\ref{eq:pivotal-mean}). In Section \ref{sec:example1} 
we present some simulations to assess the viability of the approach. In  Section
\ref{sec:example2} we apply orthogonal samples to the estimation of 
mean and variance of $L_{2}$-spectral statistics (which are often
quite complicated). 
In Section \ref{sec:test} we address the issue of testing. Specifically,
since the
orthogonal sample shares similar sampling  properties with the centralised version of the statistic, it can be
used  to estimate the finite sample distribution, and 
critical values, of the statistic under the null that
the mean of the statistic is zero.
Thus we use the orthogonal sample to estimate the distribution of the
test statistic under the null.
In Section \ref{sec:cross} we propose an
average square criterion to select the number 
of terms in the orthogonal sample. Evaluation of the orthogonal
sample requires only $O(T\log T)$ computing operations, which makes
the procedure extremely fast. 

The purpose of this paper is to propose a methodology and does not
purport to be rigourous. However, some proofs are given in the
supplementary material.

%% file: 2_estimationv2.tex
\section{Orthogonal samples and its applications}\label{sec:quad}

\subsection{Notation and assumptions}

A time series $\{X_{t}\}$ is said to be $p$th-order stationary if all  
moments up to the $p$th moment are invariant to shift
(for example, a strictly stationary
time series with finite $p$-order moment satisfies such a condition). We denote the covariance and $s$
order cumulant as  
$c(j) = \cov(X_{t},X_{t+j})$ and $\kappa_{s}(j_{1},\ldots,j_{s-1}) =
\cum(X_{t},X_{t+j_{1}},\ldots,X_{t+j_{s-1}})$. Furthermore, we define
the spectral density and $s$-order spectral density functions as 
\begin{eqnarray*}
f(\omega) = \frac{1}{2\pi}\sum_{j\in \mathbb{Z}}c(j)e^{ij\omega}
\textrm{ and }
f_{s}(\omega_{1},\ldots,\omega_{s-1}) =
\frac{1}{(2\pi)^{s}}\sum_{j_{1},\ldots,j_{s-1}\in \mathbb{Z}}\kappa_{s}(j_{1},\ldots,j_{s-1})e^{i(j_{1}\omega_{1}+\ldots
   + j_{s-1}\omega_{s-1})}. 
\end{eqnarray*}
To simplify notation we will assume that $\{X_{t}\}$ is a zero
mean time series, noting that the same methodology also works when the
mean of $\{X_{t}\}$ is constant since the DFT of a constant mean is zero at most frequencies.
We let $\Re X$ and $\Im X$ denote the real and imaginary parts of the
variable $X$.

\begin{assumption}[$p$th-order stationary and cumulant conditions]\label{assum:p}
The time series $\{X_{t}\}_{t}$ is $p$-order stationary, with $\Ex|X_{t}|^{p}<\infty$ 
and for all $2\leq s \leq p$ and  $1\leq i \leq s$
\begin{eqnarray*}
\sum_{j_{1},\ldots,j_{s-1}\in \mathbb{Z}}(1+|j_{i}|)|\kappa_{s}(j_1,\ldots,j_{s-1})|
<\infty.
\end{eqnarray*}
\end{assumption}

\subsection{Construction of orthogonal samples and pivotal statistics}

In this section the main focus will be on $A_{T}(\phi)$ (defined in (\ref{eq:ATl})).
We start by reviewing some of the well known sampling properties of $A_{T}(\phi)$.
If $\{X_{t}\}$ is a 
fourth order stationary time series which satisfies Assumption
\ref{assum:p} with $p=4$, 
then it can be shown that $A_{T}(\phi)$ is a mean squared consistent estimator of $A(\phi)$, where 
\begin{eqnarray*}
A(\phi) = \frac{1}{2\pi}\int_{0}^{2\pi}\phi(\omega)f(\omega)d\omega.
\end{eqnarray*}
Clearly, depending on the choice of $\phi$, $A_{T}(\phi)$ estimates several parameters of interest and we
give some examples below.

\begin{example}\label{example:1}
\begin{itemize}
\item[(a)] The sample autocovariance function at lag  $j$, with $\phi(\omega) =
  \exp(ij\omega_{})$, corresponds to  
\begin{eqnarray}
\label{eq:covariance}
\widehat{c}_{T}(j)=
\frac{1}{T}\sum_{k=1}^{T}\exp(ij\omega_{k})|J_{T}(\omega_{k})|^{2} =
\widetilde{c}_{T}(j) + \widetilde{c}_{T}(T-j) = \widetilde{c}_{T}(j) + O_{p}\left(\frac{|j|}{T}\right),
\end{eqnarray}
with $\widetilde{c}_{T}(j) = \frac{1}{T}\sum_{t=1}^{T-|j|}X_{t}X_{t+|j|}$.
\item[(b)]The spectral density estimator with $\phi(\omega) = b^{-1}W(\frac{\omega - \omega_{k}}{b})$.
\item[(c)] In order to test for goodness of fit of a model with
  spectral density function $g(\omega;\theta)$, \cite{p:mil-81} 
  proposed estimating the $j$th autocovariance function of the
  residuals obtained by fitting the linear model corresponding to
  $g(\omega;\theta)$ using  
\begin{eqnarray*}
\widehat{\gamma}_{T}(j)=
\frac{1}{T}\sum_{k=1}^{T}\frac{\exp(ij\omega_{k})}{g(\omega_{k};\theta)}|J_{T}(\omega_{k})|^{2}. 
\end{eqnarray*}
In this case
$\widehat{\gamma}_{T}(j)=A_{T}(e^{ij\cdot}g(\cdot;\theta)^{-1})$ and 
$\phi(\omega) = e^{ij\omega}g(\omega_{};\theta)^{-1}$.
\item[(d)] The Whittle likelihood estimator (which is asymptotically
  equivalent to the quasi-Gaussian likelihood),
where $\widehat{\theta}_{T} = \arg\min_{\theta \in \Theta}
  \mathcal{L}_{T}(\theta)$ with 
\begin{eqnarray*}
\mathcal{L}_{T}(\theta) =
\frac{1}{T}\sum_{k=1}^{T}\left(\frac{|J_{T}(\omega_{k})|^{2}}{f(\omega_{k};\theta)}
+ \log f(\omega_{k};\theta)\right)
\end{eqnarray*}
and $\Theta$ is a compact parameter space.
For the purpose of estimation and testing usually the derivative of
the likelihood is required, where 
\begin{eqnarray}
\nabla_{\theta} \mathcal{L}_{T}(\theta)  = A_{T}(\phi) +
\frac{1}{T}\sum_{k=1}^{T}
\frac{1}{f(\omega_{k};\theta)}\nabla_{\theta}f(\omega_{k};\theta)\label{eq:derL}
\end{eqnarray}
and  $\phi(\omega) = \nabla_{\theta}f(\omega_{k};\theta)^{-1}$. 
%
\end{itemize}
\end{example}
Under stationarity and some additional mixing-type and 
regularity conditions it is easily shown that $T\var[A_{T}(\phi)]= V(0) +
O(T^{-1})$, where
\begin{eqnarray}
V(0) 
&=& \frac{1}{2\pi}\int_{0}^{2\pi}f(\omega)^{2}|\left(|\phi(\omega)|^{2}
  +\phi(\omega)
\overline{\phi(-\omega)}\right)d\omega + \nonumber\\
 && \frac{1}{2\pi}\int_{0}^{2\pi}\int_{0}^{2\pi}\phi(\omega_{1})
\overline{\phi(\omega_{2})}f_{4}(\omega_{1},-\omega_{1},-\omega_{2})d\omega_{1}d\omega_{2}.\label{eq:V}
\end{eqnarray} 
It is clear that the variance has a complicated structure and cannot
be directly estimated. Instead, we
obtain an orthogonal sample associated with $A_{T}(\phi)$ to estimate $V(0)$. 
To construct the orthogonal sample  we recall
some of the pertinent features of the orthogonal sample associated with the
sample mean;   $\{\sqrt{2}\Re J_{T}(\omega_{k}), \sqrt{2}\Im J_{T}(\omega_{k});k=1,\ldots,M\}$ is a
`near uncorrelated' sequence which has similar distributional
properties as a centralised version of 
$\sqrt{T/2\pi}\bar{X}_{T} = J_{T}(0)$. Returning to 
$A_{T}(\phi)$ we observe that it is a weighted average of the
periodogram $|J_{T}(\omega_{k})|^{2}$. We now compare
$|J_{T}(\omega_{k})|^{2}$ with
$J_{T}(\omega_{k})\overline{J_{T}(\omega_{k+r})}$. Using Theorem
4.3.2, \cite{b:bri-81}, it is clear that 
$|J_{T}(\omega_{k})|^{2}$ and 
$\{\sqrt{2}\Re
J_{T}(\omega_{k})\overline{J_{T}(\omega_{k+r})},\sqrt{2}\Im
J_{T}(\omega_{k})\overline{J_{T}(\omega_{k+r})}\}$
are estimating very
different quantities (the spectral density and zero
respectively). However, they are
almost uncorrelated and in the case that
$r$ is small and $k>0$ they have approximately the same variance. This suggests
that in order to construct the orthogonal sample associated with $A_{T}(\phi)$ we replace
$|J_{T}(\omega_{k})|^{2}$ with $J_{T}(\omega_{k})\overline{J_{T}(\omega_{k+r})}$ and define
\begin{eqnarray}
\label{eq:ATr}
A_{T}(\phi;r) =
\frac{1}{T}\sum_{k=1}^{T}\phi(\omega_{k})J_{T}(\omega_{k})\overline{J_{T}(\omega_{k+r})}
\quad r\in \mathbb{Z}.
\end{eqnarray}
Note that $A_{T}(\phi;0) = A_{T}(\phi)$. In the following lemmas we show that
$\{\sqrt{2}\Re A_{T}(\phi;r), \sqrt{2}\Im A_{T}(\phi;r)\}_{r=1}^{M}$ is an orthogonal
sample to $A_{T}(\phi)$. We  first show that in general $A_{T}(\phi;0)$ and
$A_{T}(\phi;r)$ ($r>0$) have differing means.

\begin{lemma}\label{lemma:1}
Suppose that $\{X_{t}\}$ satisfies Assumption \ref{assum:p}
with $p=2$ and and $\phi(\cdot)$ is a Lipschitz
continuous bounded function. Then we have 
\begin{eqnarray*}
\Ex[A_{T}(\phi;r)] = 
\left\{
\begin{array}{cc}
\frac{1}{2\pi}\int_{0}^{2\pi}\phi(\omega)f(\omega)d\omega + O(T^{-1}) & r=0 \\
O(T^{-1}) & 0<r <T/2 \\
\end{array}
\right.
\end{eqnarray*}
\end{lemma}
{\bf PROOF} In the Supplementary material. \hfill $\Box$

\vspace{3mm}

Despite these terms having  different
expectations in the following lemma and corollary we show that they
share similar second order properties. 
\begin{theorem}\label{lemma:2}
Suppose $\{X_{t}\}$ satisfies Assumption \ref{assum:p} with $p=4$
and the function $\phi:[0,2\pi]\rightarrow \mathbb{R}$ is a Lipschitz
continuous bounded function. 
\begin{itemize}
\item[(i)]Then we have  
\begin{eqnarray*}
T\var[A_{T}(\phi)] = T\var[A_{T}(\phi;0)] = V(0) + O\left(T^{-1}\right) 
\end{eqnarray*}
\begin{eqnarray*}
T\cov[\Re A_{T}(\phi;r_{1}),\Re A_{T}(\phi;r_{2}) ] = 
\left\{
\begin{array}{cc}
\frac{1}{2}V_{}(\omega_{r}) + O(T^{-1})& 0<r_{1}=r_{2} (=r) \\
 O(T^{-1})            & 0 <  r_{1}\neq r_{2} \leq T/2 \\
\end{array}
\right.
\end{eqnarray*}
\begin{eqnarray*}
T\cov[\Im A_{T}(g;r_{1}),\Im A_{T}(\phi;r_{2}) ] = 
\left\{
\begin{array}{cc}
\frac{1}{2}V_{}(\omega_{r}) + O(T^{-1}) & 0<r_{1}=r_{2} (=r) \\
 O(T^{-1})            & 0 < r_{1}\neq r_{2} \leq T/2 \\
\end{array}
\right.
\end{eqnarray*}
and 
\begin{eqnarray*}
T\cov[\Re A_{T}(\phi;r_{1}),\Im A_{T}(\phi;r_{2}) ] = 
O(T^{-1})            \quad 0< r_{1}, r_{2} \leq T/2 
\end{eqnarray*}
\item[(ii)] Suppose, further that Assumption \ref{assum:p} holds with $p=8$,
then we have 
\begin{eqnarray}
\cov[|\sqrt{T}A_{T}(\phi;r_{1})|^{2},|\sqrt{T}A_{T}(\phi;r_{2})|^{2}]
= \left\{
\begin{array}{cc}
V(\omega_{r})^{2} + O(T^{-1}) & 0<r_{1} = r_{2} (=r) \\
O(T^{-1}) &  0 \leq r_{1} < r_{2} <T/2 \\
\end{array}
\right. \label{eq:covV2}
\end{eqnarray}
\end{itemize}
where 
\begin{eqnarray}
\label{eq:Vr}
V_{}(\omega_{r}) &=& \frac{1}{2\pi}\int_{0}^{2\pi}f(\omega)f(\omega+\omega_{r})|\left(|\phi(\omega)|^{2}
  +\phi(\omega)
\overline{\phi(-\omega-\omega_{r})}\right)d\omega + \nonumber\\
&&\frac{1}{2\pi}\int_{0}^{2\pi}\int_{0}^{2\pi}\phi(\omega_{1})
\overline{\phi(\omega_{2})}f_{4}(\omega_{1},-\omega_{1}-\omega_{r},-\omega_{2})d\omega_{1}d\omega_{2}.
\end{eqnarray} 
\end{theorem}
{\bf PROOF} In the Supplementary material. \hfill $\Box$

\vspace{3mm}

We observe that Assumption \ref{assum:p} with $p=4$
implies that the spectral density function $f(\cdot)$ and fourth order
spectral density function $f_{4}(\cdot)$
are Lipschitz continuous over each variable. These observations
immediately lead to the following result. 

\begin{corollary}\label{lemma:3}
Suppose Assumption \ref{assum:p} with $p=4$ holds and $\phi$ is
Lipschitz continuous.  Let $V(\cdot)$ be defined as in (\ref{eq:Vr}). Then we have 
\begin{eqnarray*}
\left|V(\omega_{r})  - V(0)\right| \leq K |r|T^{-1},
\end{eqnarray*}
where $K$ is a finite constant that does not depend on $r$ or $T$. 
\end{corollary}
Theorem \ref{lemma:2}(i) and Lemma \ref{lemma:3} together imply for $M<<T$,
that the sequence 
\\
$\{\sqrt{2}\Re A_{T}(\phi;r), \sqrt{2}\Im
A_{T}(\phi;r);r=1,\ldots,M \}$ are `near uncorrelated' random
variables with approximately the same variance, $V(0)$.  Based 
on these 
observations we propose the following estimator of $V(0)$
\begin{eqnarray}
\label{eq:estV}
\widehat{V}_{M}(0) = \frac{T}{2M}\sum_{r=1}^{M}\left(2|\Re
  A_{T}(\phi;r)|^{2} + 2|\Im A_{T}(\phi;r)|^{2}\right) = \frac{T}{M}\sum_{r=1}^{M}|A_{T}(\phi;r)|^{2}.
\end{eqnarray}
Below we obtain the orthogonal samples associated with the estimators
described in Example \ref{example:1}.
\begin{example}\label{example:3}
\begin{itemize}
\item[(a)] We recall that the sample covariance is
\begin{eqnarray*}
A_{T}(e^{ij\cdot}) = \frac{1}{T}\sum_{k=1}^{T}|J_{T}(\omega_{k})|^{2}e^{ij\omega_{k}}\approx \frac{1}{T}\sum_{t=1}^{T-j}X_{t}X_{t+j}
\end{eqnarray*}
and the orthogonal sample is approximately
\begin{eqnarray*}
A_{T}(e^{ij\cdot};r) =
\frac{1}{T}\sum_{k=1}^{T}|J_{T}(\omega_{k})|^{2}e^{ij\omega_{k}}
\approx \frac{1}{T}\sum_{t=1}^{T-j}X_{t}X_{t+j}e^{-it\omega_{r}}.
\end{eqnarray*}
Thus the sample covariance is a sample mean,  and, as expected, the
orthogonal sample is analogous to the DFT $J_{T}(\omega_{k})$, but with
$X_{t}X_{t+j}$ replacing $X_{t}$.
\item In general, unlike the sample covariance given above, 
$A_{T}(\phi)$ will not be a ``mean-like'',   
but a quadratic form $A_{T}(\phi) =
  \frac{1}{T}\sum_{t,\tau=1}^{T}\Phi_{T}(t-\tau)X_{t}X_{\tau}$, where
  $\Phi_{T}(t-\tau) =
  \frac{1}{T}\sum_{k=1}^{T}\phi(\omega_{k})e^{i(t-\tau)\omega_k}$. 
Straightforward calculations how that the corresponding  
  orthogonal sample can be written as $A_{T}(\phi;r) =
  \frac{1}{T}\sum_{t,\tau=1}^{T}\Phi_{T}(t-\tau)X_{t}X_{\tau}e^{-i\tau\omega_{r}}$.
\item[(b)] The orthogonal sample for the spectral density estimator is 
\begin{eqnarray*}
\widehat{f}(\omega;r) = \frac{1}{bT}\sum_{k=1}^{T}W\left(\frac{\omega -\omega_{k}}{b}\right)J_{T}(\omega_{k})\overline{J_{T}(\omega_{k+r})}.
\end{eqnarray*}
Note that since $\phi = \frac{1}{b}W\left(\frac{\omega
    -\omega_{k}}{b}\right)$ is not a bounded function (over $b$) the 
rates in Lemma \ref{lemma:1} and Theorem \ref{lemma:2} do not 
hold, and some adjustment of the rate is necessary.
\item[(c)] The orthogonal sample for $\widehat{\gamma}_{T}(j)$ is 
$A_{T}( e^{ij\cdot}g(\cdot;\theta)^{-1};r)=T^{-1}\sum_{k=1}^{T} e^{ij\omega_{k}}g(\omega_{k};\theta)^{-1}J_{T}(\omega_{k})\overline{J_{T}(\omega_{k+r})}$.
\item[(d)] If the aim to test $H_{0}:\theta = \theta_0$ versus
  $H_{A}:\theta \neq \theta_0$ using the score test based on 
the Whittle likelihood,
$\sqrt{T}\nabla_{\theta_0}\mathcal{L}_{T}(\theta)$,  
then we require an estimator of the variance 
\begin{eqnarray*}
V = \lim_{T\rightarrow \infty}\var[\sqrt{T}\nabla_{\theta_0}
\mathcal{L}_{T}(\theta) ] = \lim_{T\rightarrow
  \infty}\var[\sqrt{T}A_{T}(\nabla_{\theta_0}f(\omega_{k};\theta)^{-1})]. 
\end{eqnarray*}
In this case the orthogonal sample is
\\*
$A_{T}(\nabla_{\theta_0}f(\omega_{k};\theta)^{-1};r) =
T^{-1}\sum_{k=1}^{T}\nabla_{\theta_0}f(\omega_{k};\theta)^{-1}
J_{T}(\omega_{k})\overline{J_{T}(\omega_{k+r})}$.
\end{itemize}
\end{example}

We return to the Whittle likelihood considered in Example
\ref{example:1}(d).  In the above example, we use the Whittle
likelihood for hypothesis testing. However, the Whittle likelihood is
also used in estimation, where $\widehat{\theta}_{T}=
\arg\min \mathcal{L}_{T}(\theta)$ is an estimator of the
true parameter, $\theta_0$.  By using (\ref{eq:derL}) and the
Taylor series expansion it is well known that 
\begin{eqnarray*}
\sqrt{T}\left(\widehat{\theta}_{T} - \theta_0\right) \Dcon 
\mathcal{N}\left(0,W_{\theta_0}^{-1}V_{\theta_0}W_{\theta_0}^{-1}\right),
\end{eqnarray*}
where $W_{\theta_0}=\frac{1}{2\pi}\int_{0}^{2\pi}f(\omega)^{-2}\nabla_{\theta_0}
f_{\theta}(\omega;\theta)\nabla_{\theta_0}
f_{\theta}(\omega;\theta)^{\prime}d\omega$ and $V_{\theta_0} = \lim_{T\rightarrow
  \infty}T\var[A_{T}(\nabla_{\theta_0}f(\omega_{k};\theta)^{-1})]$.
However, $\theta_0$ is unknown, and we only have an estimator 
$\widehat{\theta}_{T}$. Instead, we replace $\theta_0$  with $\widehat{\theta}_{T}$
and use 
$$A_{T}(\nabla_{\widehat{\theta}_{T}}
f(\omega_{k};\widehat{\theta})^{-1};r) = \frac{1}{T}\sum_{k=1}^{T}\nabla_{\widehat{\theta}_{T}}
f(\omega_{k};\widehat{\theta})^{-1}J_{T}(\omega_{k})\overline{J_{T}(\omega_{k+r})}$$
as the orthogonal sample and 
\begin{eqnarray}
\label{eq:Vwhittle}
\widehat{V}_{\widehat{\theta},M}(0) = \frac{1}{M}\sum_{r=1}^{M}
\left|\frac{1}{\sqrt{T}}\sum_{k=1}^{T}\nabla_{\theta}
f(\omega_{k};\widehat{\theta})^{-1}J_{T}(\omega_{k})\overline{J_{T}(\omega_{k+r})}\right|^{2}
\end{eqnarray}
as an estimator of $V_{\theta}(0)=\lim_{T\rightarrow \infty}T\var[A_{T}(\nabla_{\theta}f(\cdot;\theta)^{-1})]$.

It is straightforward to generalize this idea to estimate $V_{\theta}(0) = \var\left[ \sqrt{T}A_{T}(\phi_{\theta})\right]$, where
$\theta_0$ is unknown and only an estimator $\widehat{\theta}_{T}$ of $\theta_0$ is observed. We
suggest using  
\begin{eqnarray}
\label{eq:varA}
\widehat{V}_{\widehat{\theta},M}(0)=
\frac{T}{M}\sum_{r=1}^{M}|A_{T}(\phi_{\widehat{\theta}};r)|^{2},
\textrm{ where }\quad
A_{T}(\phi_{\widehat{\theta}};r) =\frac{1}{T}\sum_{k=1}^{T}\phi(\omega_{k};\widehat{\theta})J_{T}(\omega_{k})\overline{J_{T}(\omega_{k+r})}
\end{eqnarray}
as an estimator of $V_{\theta} = \lim_{T\rightarrow \infty}T\var[A_{T}(\phi_{\theta})^{}]$.

In Lemmas \ref{lemma:4} and \ref{lemma:composite}, below, we show that
$\widehat{V}_{M}(0)$ and $\widehat{V}_{\widehat{\theta},M}(0)$ are
consistent estimators of the variance. But, in most applications
variance estimation is mainly required in the construction of  confidence intervals or
test hypothesis. In which case the main object of interest is the studentized statistic 
\begin{eqnarray}
\label{eq:TMM}
T_{M} = \frac{\sqrt{T}[A_{T}(\phi) - A(\phi)]}{\sqrt{\widehat{V}_{M}(0)}}. 
\end{eqnarray} 
Since $\widehat{V}_{M}(0)$, is a consistent estimator of $V(0)$ as
$M/T\rightarrow \infty$ with $M,T\rightarrow \infty$, under suitable conditions we would expect 
$T_{M}\Dcon N(0,1)$. However, such an approximation does not take into
account that $\widehat{V}_{M}(0)$ is an estimator of the variance. We want
a better finite sample approximation that takes into account that $M$
is fixed. This requires the following theorem. 

\begin{theorem}
Let us suppose that $\{X_{t}\}$ is a stationary $\alpha$-mixing time
series, where the $\alpha$-mixing coefficient $\alpha(t)$ is such that
$\alpha(t)\leq K|t|^{-s}$ (for $|t|\neq 0$), where $s>6$ and
$K<\infty$ and for some $r > 4s/(s-6)$ we have
$\Ex|X_{t}|^{r}<\infty$.

Let $A_{T}(\phi)$ and $A_{T}(\phi;r)$ be defined as in (\ref{eq:ATl}) and
(\ref{eq:ATr}) respectively. We 
assume that $\phi:[0,2\pi]\rightarrow \mathbb{R}$ has a
bounded second derivative and $A_{T}(\phi)$ is a real-valued random
variable. 
Let $A_{M,T} =(A_{T}(\phi;1),\ldots,A_{T}(\phi;M))^{\prime}$. Then for $M$ fixed
we have 
\begin{eqnarray}
\label{eq:CLT}
\sqrt{\frac{T}{V(0)}}
\left(
\begin{array}{c}
A_{T}(\phi)-A(\phi) \\
\sqrt{2}\Re A_{M,T} \\
\sqrt{2}\Im A_{M,T} \\
\end{array}
\right) \Dcon 
\mathcal{N}\left(0,I_{2M+1}\right), 
\end{eqnarray}
where $I_{2M+1}$ denotes the identity matrix of dimension $2M+1$.
\end{theorem}
{\bf PROOF} The proof immediately follows from Theorem 2.1,
\cite{p:lee-sub-10}. 

Note that the stated conditions imply that Assumption
\ref{assum:p} holds with $p=4$, see \cite{p:sta-88}, Theorem 3, part
(2) and Remark 3.1, \cite{p:neu-96}.
\hfill $\Box$

\vspace{3mm}

Using the asymptotic
independence of $\Re A_{M,T}$ and $\Im A_{M,T}$ (proved in the above
theorem) we observe for fixed $M$
\begin{eqnarray*}
M\frac{\widehat{V}_{M}(0)}{V(0)} \Dcon \chi^{2}_{2M} 
\end{eqnarray*}
as $T\rightarrow \infty$, where $\chi^{2}_{2M}$ denotes a chi-square
distribution with $2M$ degrees of freedom. Furthermore, since
$A_{T}(\phi)$ is independent of $\Re A_{M,T}$ and $\Im A_{M,T}$ 
for fixed $M$ we have 
\begin{eqnarray}
\label{eq:TM}
T_{M}\Dcon t_{2M}
\end{eqnarray}
as $T\rightarrow \infty$, where $T_{M}$ is defined in (\ref{eq:TMM})
and 
$t_{2M}$ denotes the $t$-distribution
with $2M$ degrees of freedom. 

\begin{remark}\label{remark:covariance}
The above method can be generalized to estimate the covariance between
several estimators which take the form (\ref{eq:ATl}). Let 
${\boldsymbol A} = (A(\phi_{1}),\ldots,A(\phi_{p}))$ denote a $p$-dimensional
parameter and
${\boldsymbol A}_{T} = (A_{T}(\phi_{1}),\ldots,A_{T}(\phi_{p}))$ their corresponding
estimators. Further, let ${\boldsymbol A}_{T}(r) =(A_{T}(\phi_{1};r),\ldots,A_{T}(\phi_{p};r))$
denote the orthogonal sample vector associated with ${\boldsymbol
  A}_{T}$. It can be shown that $\var[\sqrt{T}{\boldsymbol A}_{T}] =
\Sigma +O(T^{-1})$ where
\begin{eqnarray}
\Sigma_{j_1,j_2} &=&  \frac{1}{2\pi}\int_{0}^{2\pi}f(\omega)^{2}
\left[\phi_{j_{1}}(\omega)\overline{\phi_{j_{2}}(\omega)} +\phi_{j_{1}}(\omega)\overline{\phi_{j_{2}}(-\omega)}\right]
d\omega + \nonumber\\
&&\frac{1}{2\pi}\int_{0}^{2\pi}\int_{0}^{2\pi}\phi_{j_{1}}(\omega_{1})\overline{\phi_{j_{2}}(\omega_{2})}
f_{4}(\omega_{1},-\omega_{1},-\omega_{2})d\omega_{1}d\omega_{2}. \label{eq:sigmaj1j2}
\end{eqnarray}
Based on similar ideas to those presented above we can estimate the
variance $\Sigma$ with 
\begin{eqnarray*}
\widehat{\Sigma}_{M} = \frac{T}{M}\sum_{r=1}^{M}\left(\Re {\boldsymbol
    A}_{T}(r) {\boldsymbol A}_{T}(r)^{*} + \Im {\boldsymbol
    A}_{T}(r) {\boldsymbol A}_{T}(r)^{*}\right)
\end{eqnarray*}
where $ {\boldsymbol A}_{T}(r)^{*}$ denote the complex conjugate and
transpose of ${\boldsymbol A}_{T}(r)$. 
Furthermore the statistic 
\begin{eqnarray*}
T\left({\boldsymbol A}_{T} - {\boldsymbol
    A}\right)^{\prime}\widehat{\Sigma}_{M}^{-1}\left({\boldsymbol
    A}_{T} - {\boldsymbol A}\right)\Dcon T_{p,2M}^{2},
\end{eqnarray*}
where $T^{2}_{p,2M}$ denotes 
Hotelling's T-squared distribution with $2M$-degrees of freedom.
\end{remark}

Finally, we show that $\widehat{V}_{M}(0)$ is a mean square
consistent estimator of $V$.
 \begin{lemma}\label{lemma:4}
Suppose Assumption \ref{assum:p} with $p=8$ is satisfied and $\phi$ is
Lipschitz continuous. Let $\widehat{V}_{M}(0)$ be defined as in (\ref{eq:estV}).
Then we have $|\Ex[\widehat{V}_{M}(0)] - V(0)| = O(M/T)$ and 
\begin{eqnarray}
\label{eq:mse}
\Ex\left(\widehat{V}_{M}(0) - V(0) \right)^{2} = O\left(
  \frac{M^{2}}{T^{2}} + \frac{1}{M} \right).
\end{eqnarray}
\end{lemma}
{\bf PROOF} In the Supplementary material. \hfill $\Box$

\vspace{3mm}
It is interesting to note that the estimator of $\widehat{V}_{M}(0)$ is analogous to kernel
estimators in nonparametric regression, where $M$ plays the role of
window width (bandwidth multiplied by the length of time
series). From Lemma \ref{lemma:4} we observe if $M$ is large,
then $\widehat{V}_{M}(0)$ can have a large bias. On the other hand, if
$M$ is small the bias is small but the variance is large. 
However, by using a small
$M$, we can correct for the large variance by using 
the $t$-distribution approximation given in (\ref{eq:TM}). The only
real cost of using small $M$ are slightly larger critical values (due
to using a t-distribution with a small number of degrees of freedom). 

Using the above we show that $\widehat{V}_{\widehat{\theta},M}(0)$ consistently estimates $V_{\theta}$. 
\begin{lemma}\label{lemma:composite}
Suppose Assumption \ref{assum:p} with $p=8$ is satisfied. Let $\widehat{\theta}_{T}$ be a consistent
estimator of $\theta$ such that
 $|\widehat{\theta}_{T} - \theta| =
O_{p}(T^{-1/2})$, $\sup_{\theta,\omega}|\frac{\partial \phi(\omega;\theta)}{\partial
  \omega}|<\infty$ and $\sup_{\theta,\omega}|\frac{\partial^{2} \phi(\omega;\theta)}{\partial
  \theta^{2}}|<\infty$. Let $\widehat{V}_{\widehat{\theta},M}(0)$ be
defined as in (\ref{eq:varA}). Then we have 
\begin{eqnarray*}
\left|\widehat{V}_{\widehat{\theta},M}(0) - V_{\theta}\right| = O_{p}\left(
  \frac{M^{}}{T^{}} + \frac{1}{\sqrt{M}} \right),
\end{eqnarray*}
where $V_{\theta} = \lim_{T\rightarrow \infty}T\var[A_{T}(\phi_{\theta})]$.
\end{lemma}
{\bf PROOF} In the Supplementary material. \hfill $\Box$
\vspace{3mm}

\begin{example}
We apply the above result to show consistency of the variance
estimator in (\ref{eq:Vwhittle}) (corresponding to the Whittle
likelihood estimator). We observe that if $f(\omega;\theta)$
is uniformly bounded away from zero and uniformly bounded from above  
for all $\theta\in \Theta$ and $\omega\in [0,2\pi]$, and its first and second derivatives with respect to $\theta$
and $\omega$ are uniformly bounded, then (\ref{eq:Vwhittle}) is a consistent estimator of $V_{\theta}$
if $M/T\rightarrow 0$ as $M\rightarrow \infty$ and $T\rightarrow
\infty$. 
\end{example}

\subsection{Example 1}\label{sec:example1}

\begin{figure}
\centering
\includegraphics[scale=0.35]{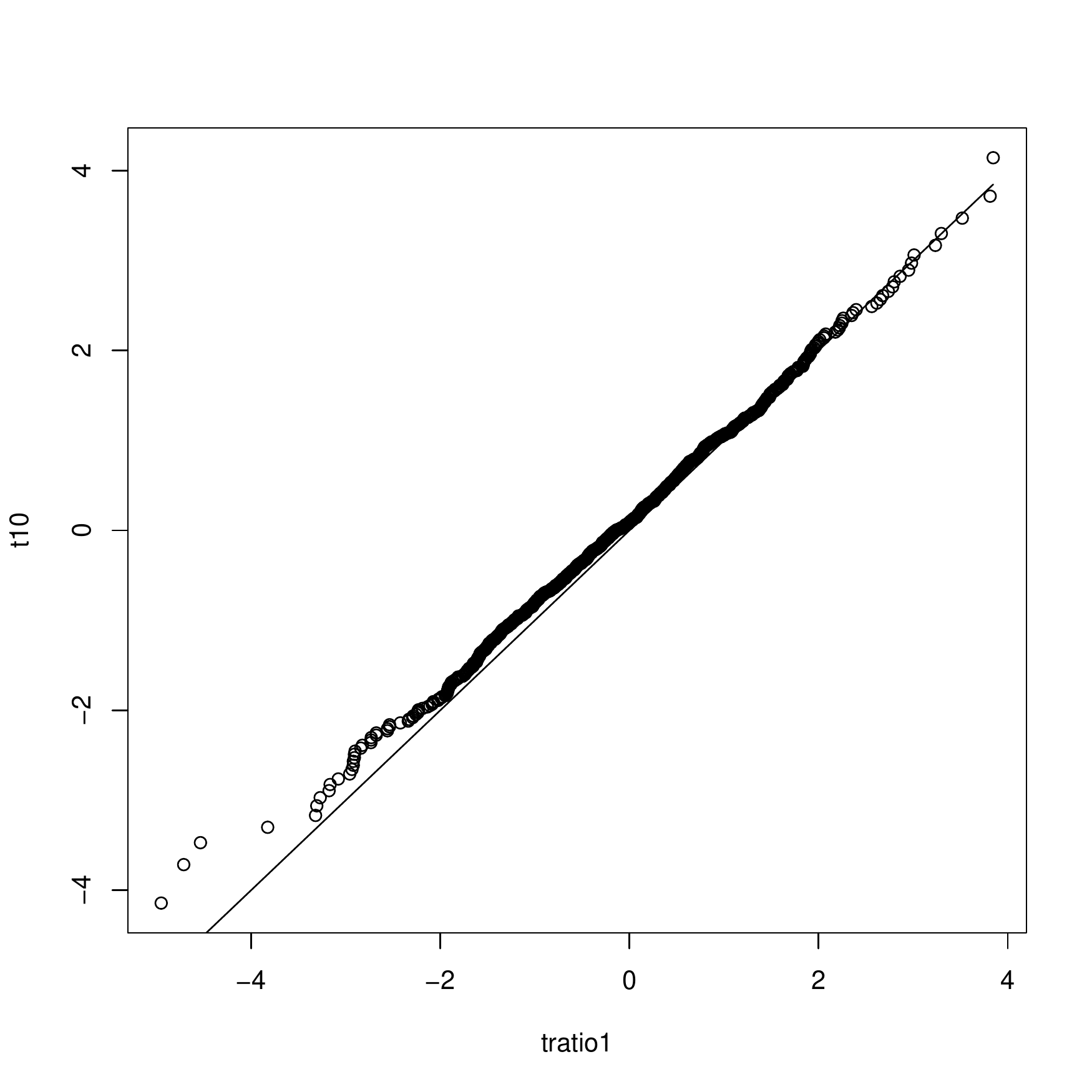}
\includegraphics[scale=0.35]{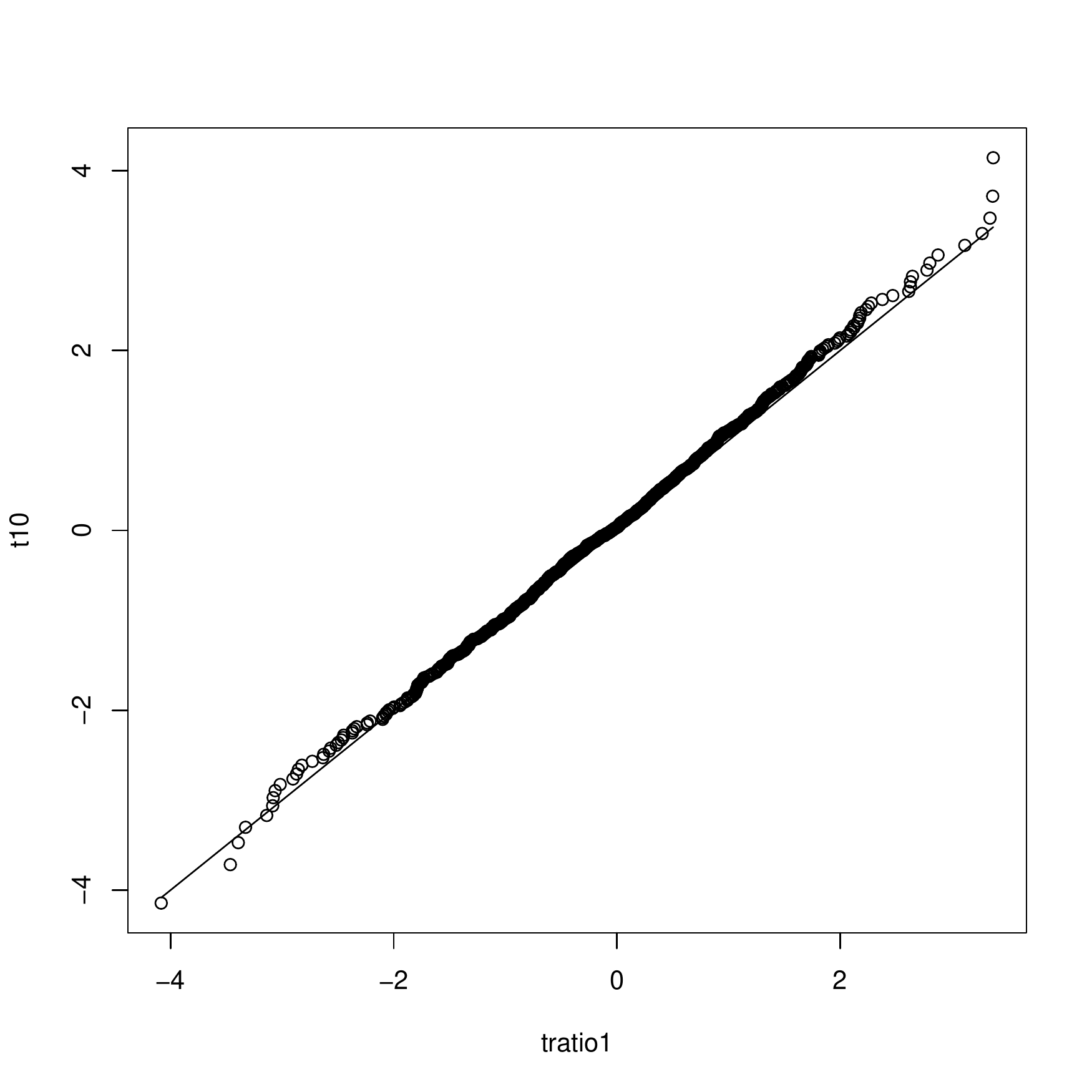}
\caption{$X_{t}$ are iid standard normal random variables. Left:
  $T=100$. Right $T=200$. 
\label{fig:QQ1}}
\end{figure}

\begin{figure}
\centering
\includegraphics[scale=0.35]{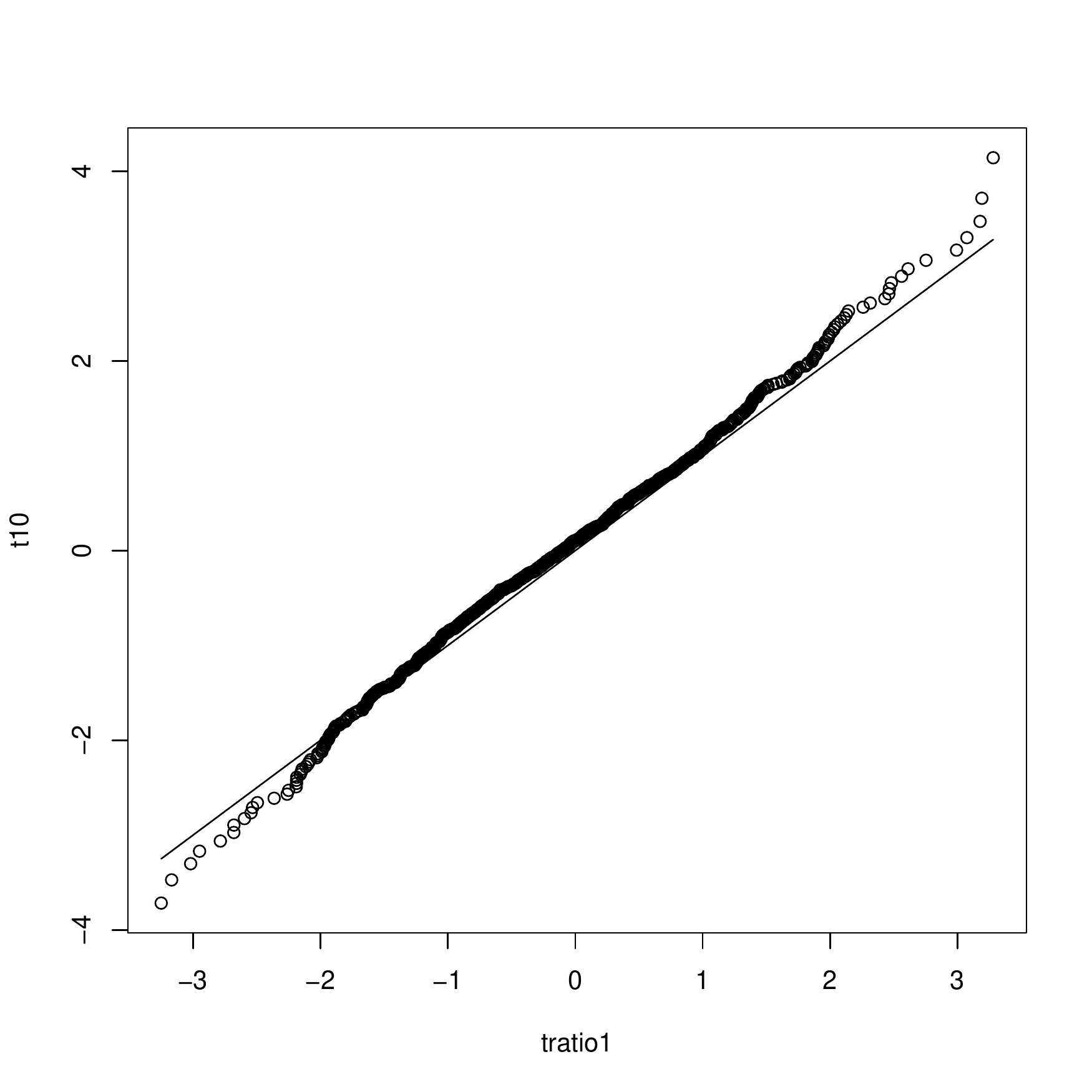}
\includegraphics[scale=0.35]{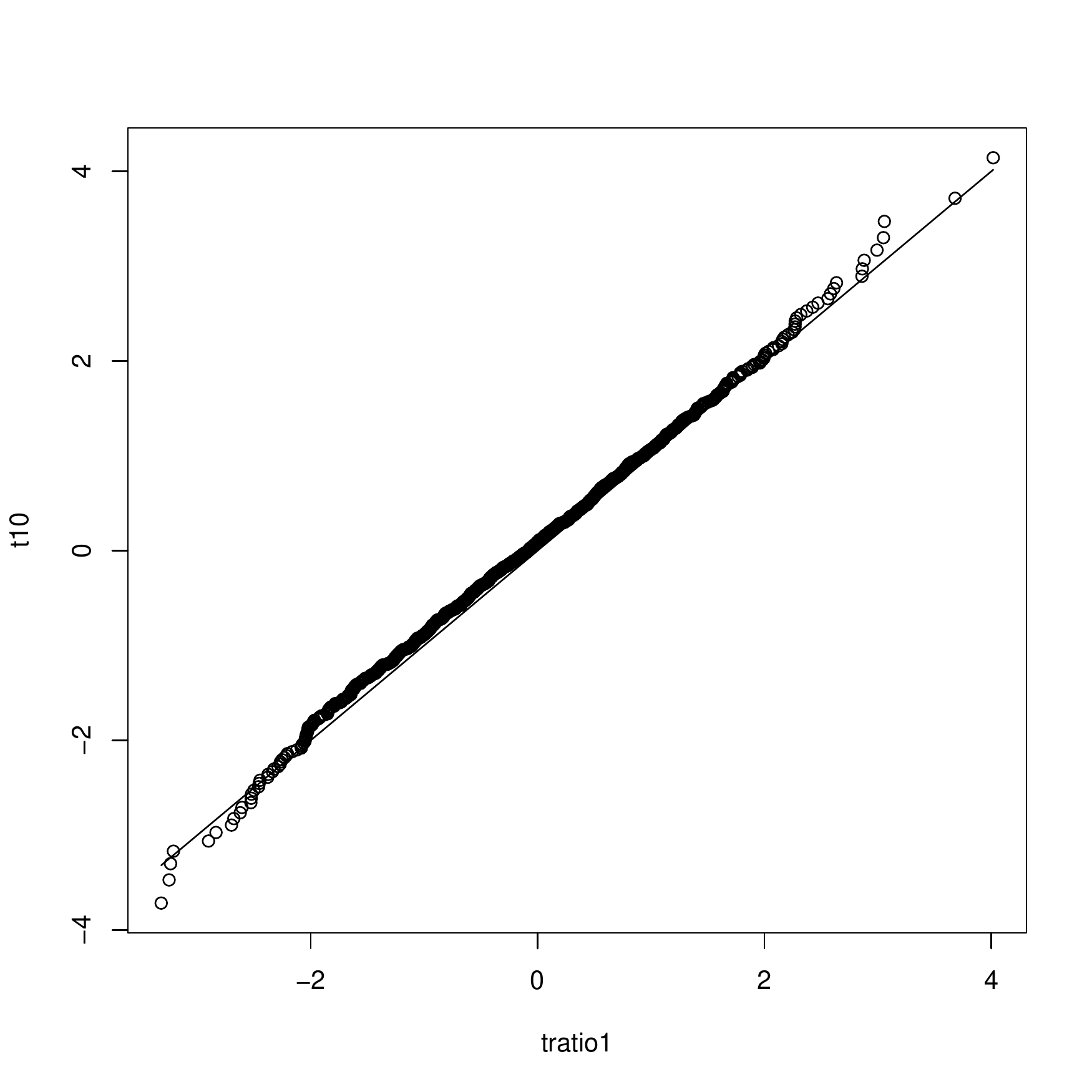}
\caption{$X_{t}$ satisfies (\ref{eq:Xtuncorrelated}) where the
  innovations are from t-distribution with 5df. 
 Left:
  $T=100$. Right $T=200$. 
\label{fig:QQ2}}
\end{figure}

\begin{figure}
\centering
\includegraphics[scale=0.35]{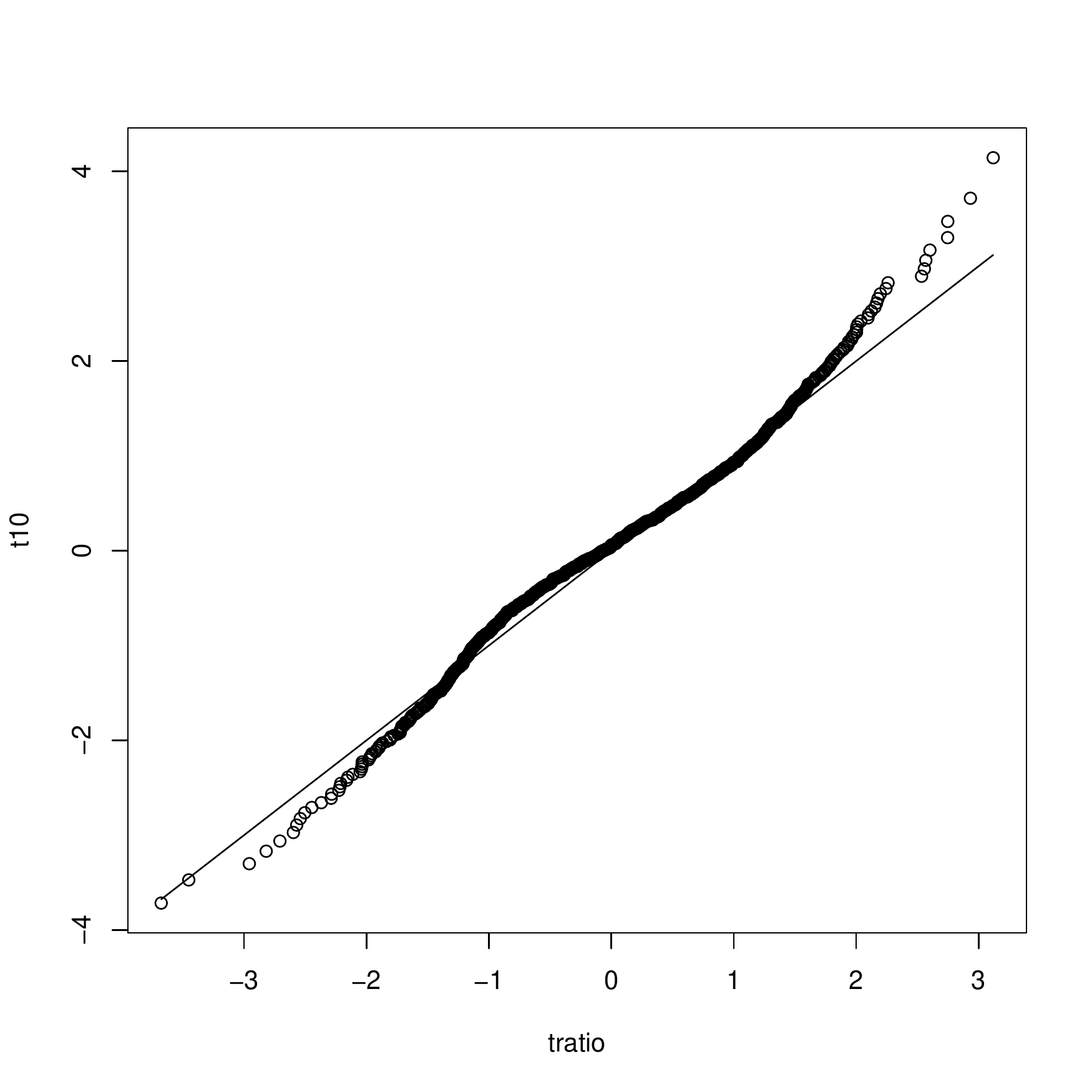}
\includegraphics[scale=0.35]{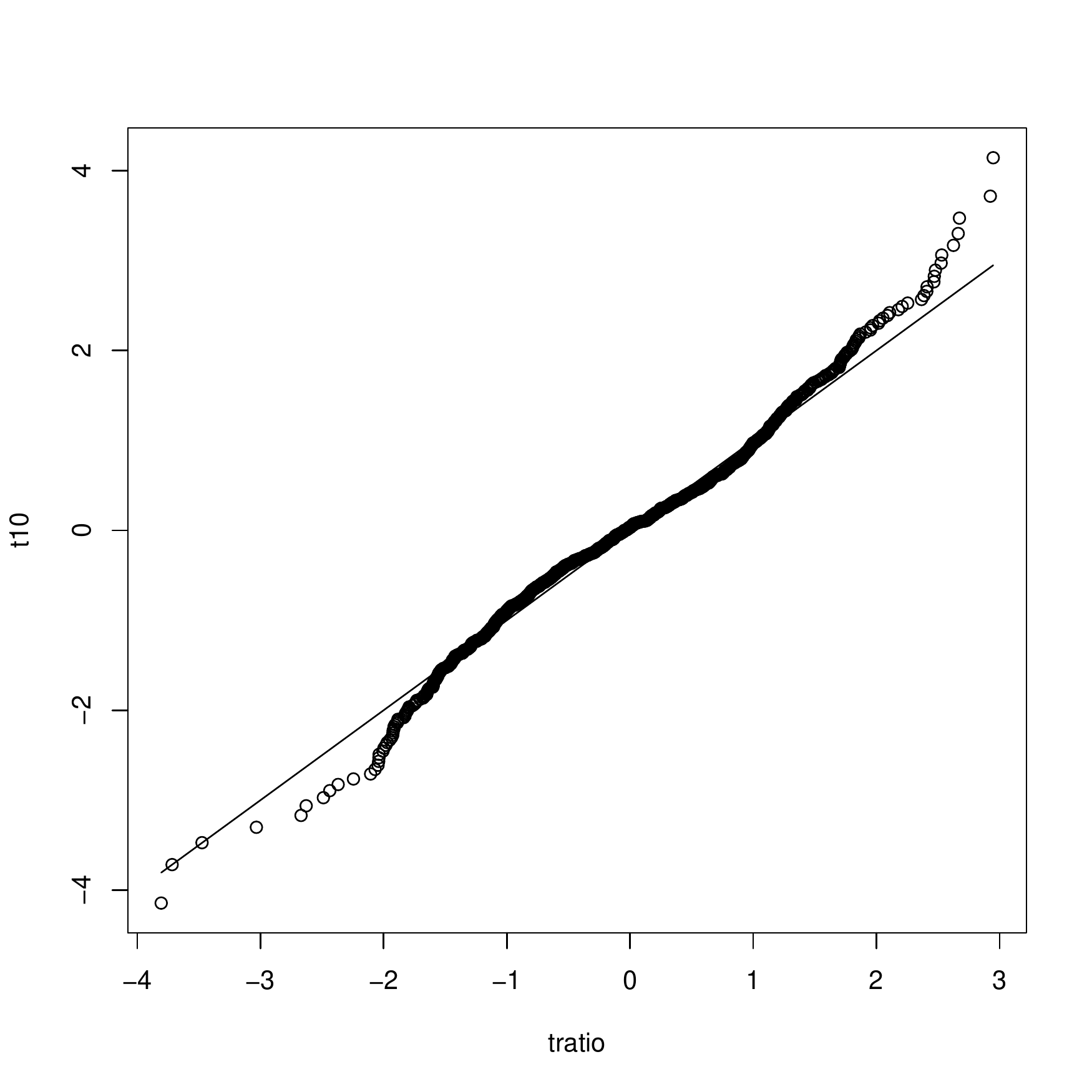}
\caption{$X_{t}$ satisfies (\ref{eq:Xtuncorrelated}) where the
  innovations are an ARCH process (the fourth moment does
  not exist). Left:
  $T=100$. Right $T=200$. 
\label{fig:QQ3}}
\end{figure}

We illustrate the result in (\ref{eq:TM}) with some
simulations. Let $\widehat{c}_{T}(1)=A_{T}(e^{i\cdot})$ denote the estimator of the
covariance at lag one (defined in (\ref{eq:covariance})) and
$\{A_{T}(e^{i\cdot};r)\}$ the corresponding orthogonal sample. We use
$M=5$ and define the studentized statistic
 \begin{eqnarray}
\label{eq:T10}
T_{10}= \frac{A_{T}(e^{i\cdot})-c(1)}{\sqrt{\frac{1}{5}\sum_{r=1}^{5}|A_{T}(e^{i\cdot};r)|^{2}}}.
\end{eqnarray}
We focus on  models where there is no correlation, thus $c(j)=0$
for $j\neq 0$, but possible higher order dependence. 
Let $\{X_{t}\}$ be an uncorrelated time series defined by 
\begin{eqnarray}
\label{eq:Xtuncorrelated}
X_{t} = \sum_{j=0}^{\infty}0.6^{j}\varepsilon_{t-j} -
\frac{0.6}{1-0.6^{2}}\varepsilon_{t+1}
\end{eqnarray}
where $\{\varepsilon_{t}\}$ are uncorrelated random variables. 
The models we consider are 
\begin{itemize}
\item[(i)] $\{X_{t}\}$ are independent, identically distributed (iid)
  normal random variables 
\item[(ii)]
$\{X_{t}\}$ satisfies (\ref{eq:Xtuncorrelated}) where the innovations $\varepsilon_{t}$ are iid t-distributed random
variables with 5df (thus $\{X_{t}\}$ is an uncorrelated, non-causal linear time
series with a finite fourth moment) 
\item[(iii)] $\{X_{t}\}$ satisfies
(\ref{eq:Xtuncorrelated}) where the innovations $\varepsilon_{t}$
satisfy the ARCH$(1)$ representation 
$\varepsilon_{t} = \sigma_{t}Z_{t}$ with $\sigma_{t}^{2} = 1
+0.7\varepsilon_{t-1}^{2}$ and $\{Z_{t}\}$ are Gaussian random
variables. $\{X_{t}\}$ is a nonlinear, uncorrelated time series whose
fourth moment is not finite, thus $A_{T}(e^{i\cdot})$ will not have a
finite variance.  
\end{itemize}
For each model a time series of size
$T=100$ and $200$ is generated and $T_{10}$ evaluated (see equation
(\ref{eq:T10})). This is done over $1000$ replications. The
QQplot of $T_{10}$ against the quantiles of a t-distribution with 10
df are given in Figures \ref{fig:QQ1} (for model (i))  \ref{fig:QQ2}
(for model (ii)) and \ref{fig:QQ3} (for model (iii)). It is reassuring to see 
that even when the sample size is relatively small ($T=100$), 
for model (i) and (ii), the finite sample
distribution of $T_{10}$ is close to $t_{10}$. Furthemore, the small
deviation seen in the tails when $T=100$ is reduced when the sample
size is increased to $T=200$.  For model (iii) $\Ex[X_{t}^{4}]$ is not
finite,  thus $\Ex|A_{T}(e^{i\cdot};r)|^{2}$ is not finite and the assumptions
which underpin (\ref{eq:TM}) do not hold. This is apparent in 
Figure \ref{fig:QQ3}, where the  $t$-distribution seems
inappropriate. It
is interesting to investigate what the distribution of $T_{10}$ is in
this case and we leave this for future research.

\subsection{Example 2}\label{sec:example2}

Suppose that $\{X_{t}\}$ and $\{Y_{t}\}$ are two time series which are
jointly stationary and with univariate 
spectral densities $f_{X}$ and $f_{Y}$ respectively. 
We now apply the above methodology to testing for equality of spectral
densities i.e. $H_{0}:f_{X}(\omega) = f_{Y}(\omega)$ for all $\omega
\in [0,2\pi]$ against $H_{A}:f_{X}(\omega) \neq f_{Y}(\omega)$ for
some $\omega$ (with non-zero measure). 
\cite{p:eic-08} and \cite{p:det-pap-09} propose testing for equality of
the spectral densities using an $L_{2}$-distance, this requires
estimators for $f_{X}$ and $f_{Y}$. Define
\begin{eqnarray*}
\widehat{f}_{X}(\omega_l;r) &=&
\frac{1}{bT}\sum_{k=1}^{T}W\left(\frac{\omega_l - \omega_k}{b}\right)
J_{X,T}(\omega_k)\overline{J_{X,T}(\omega_{k+r})} \\
\widehat{f}_{Y}(\omega_l;r) &=&
\frac{1}{bT}\sum_{k=1}^{T}W\left(\frac{\omega_l - \omega_k}{b}\right)
J_{Y,T}(\omega_k)\overline{J_{Y,T}(\omega_{k+r})},
\end{eqnarray*}
where $J_{X,T}(\omega_k)$ and $J_{Y,T}(\omega_k)$ denote the DFT of 
$\{X_{t}\}_{t=1}^{T}$ and $\{Y_{t}\}_{t=1}^{T}$ respectively and
$W(\cdot)$ is a spectral window. 
It is clear that $\widehat{f}_{X}(\omega_{l}) =
\widehat{f}_{X}(\omega_l;0)$ and 
$\widehat{f}_{Y}(\omega_{l})=\widehat{f}_{Y}(\omega_l;0)$
are estimators of the spectral density and, from Example \ref{example:3}(b),
$\widehat{f}_{X}(\omega_l;r)$ and $\widehat{f}_{Y}(\omega_l;r)$
($r\neq 0$) are the corresponding orthogonal sample. 

An obvious method for testing equality of the spectral densities is to use the $L_{2}$-statistic
\begin{eqnarray}
\label{eq:ST1}
S_{T} = \frac{2}{T}\sum_{j=1}^{T/2}\left|\widehat{f}_{X}(\omega_{j})-\widehat{f}_{Y}(\omega_{j})\right|^{2},
\end{eqnarray}
where $\widehat{f}_{T}(\omega)$ and $\widehat{f}_{T}(\omega)$ are
estimators of the spectral density function. 
Let $\mu_{T}$ and $\sigma^{2}_{T}$ denote the mean and
variance of $S_{T}$ under the null hypothesis. 
Under the null hypothesis and suitable mixing conditions
it can be shown that
\begin{eqnarray}
\label{eq:AsymptoticNormal}
\frac{S_{T}-\mu_{T}}{\sigma_{T}} \Dcon
\mathcal{N}(0,1) \quad \textrm{as }T\rightarrow \infty. 
\end{eqnarray}
Expressions for $\mu_{T}$ and $\sigma^{2}_{T}$ can be
deduced from \cite{p:eic-08}, Theorem 3.11. However, these expression
are rather complicated.
Alternatively, a rather painless method is to use the orthogonal
sample to estimate the mean and variance. The critical insight, is that 
under the null hypothesis 
\begin{eqnarray*}
\Ex[\widehat{f}_{X}(\omega_{k})-\widehat{f}_{Y}(\omega_{k})] = 
0.
\end{eqnarray*} 
However, regardless of whether the null holds or not, 
$\Ex[\widehat{f}_{X}(\omega_{k};r)]=O(T^{-1})$ and 
$\Ex[\widehat{f}_{Y}(\omega_{k};r)]=O(T^{-1})$, thus 
\begin{eqnarray*}
\Ex[\widehat{f}_{X}(\omega_{k};r)-\widehat{f}_{Y}(\omega_{k};r)]\approx
0.
\end{eqnarray*}
Therefore under the null both
$\widehat{f}_{X}(\omega_{k})-\widehat{f}_{Y}(\omega_{k})$ and
$\widehat{f}_{X}(\omega_{k};r)-\widehat{f}_{Y}(\omega_{k};r)$  share (approximately)
the same mean. Furthermore,
$\widehat{f}_{X}(\omega_{k})-\widehat{f}_{Y}(\omega_{k})$,
$\sqrt{2}\Re[\widehat{f}_{X}(\omega_{k};r)-\widehat{f}_{Y}(\omega_{k};r)]$ and 
$\sqrt{2}\Im[\widehat{f}_{X}(\omega_{k};r)-\widehat{f}_{Y}(\omega_{k};r)]$ 
asymptotically have the same variance. We now build the orthogonal
sample associated with $S_{T}$. Let 
\begin{eqnarray*}
S_{R,T}(r) &=&
\frac{4}{T}\sum_{j=1}^{T/2}|\Re\widehat{f}_{X}(\omega_{j};r)-\Re\widehat{f}_{Y}(\omega_{j};r)|^{2} \\
\textrm{ and }S_{I,T}(r) &=& \frac{4}{T}\sum_{j=1}^{T/2}|\Im\widehat{f}_{X}(\omega_{j};r)-\Im\widehat{f}_{Y}(\omega_{j};r)|^{2}.
\end{eqnarray*}
Tedious calculations show that if $M<<T$, then
$\{S_{R,T}(r),S_{I,T}(r)\}_{r=1}^{M}$ have asymptotically the same mean and
variance. Furthermore, if the null is true then the mean and variance of
$\{S_{R,T}(r),S_{I,T}(r)\}_{r=1}^{M}$ and $S_{T}$ are asymptotically the same. 
Let 
\begin{eqnarray*}
\boldsymbol{S}_{T,M} =
(S_{T},S_{R,T}(1),S_{I,T}(1),\ldots,S_{R,T}(M),S_{R,I}(M)),
\end{eqnarray*} 
then under sufficient mixing conditions it can be shown that
\begin{eqnarray}
\label{eq:sigmaT}
\sigma_{T}^{-1}\left( \boldsymbol{S}_{T,M} - \mu_{T}\boldsymbol{1}
\right)\Dcon \mathcal{N}\left(0,I_{2M+1}\right), 
\end{eqnarray}
as $b\rightarrow 0$, $bT\rightarrow \infty$ and $T\rightarrow \infty$,
where $\boldsymbol{1}$ is a $(2M+1)$-dimensional vector of ones.
We estimate the mean and variance $\mu_{T}$ and $\sigma^{2}_{T}$
using the sample mean and  variance of the orthogonal sample $\{S_{R,T}(r),S_{I,T}(r)\}_{r=1}^{M}$. In
particular, we have 
\begin{eqnarray}
\widehat{\mu}_{T} &=& \frac{1}{2M}\sum_{r=1}^{M}[S_{R,T}(r)+S_{I,T}(r)] \nonumber\\
\textrm{ and } 
\widehat{\sigma}_{T}^{2} &=&  
  \frac{1}{2M}\sum_{r=1}^{M}\left[\left( S_{R,T}(r) -
      \widehat{\mu}_{T}\right)^{2} + \left( S_{I,T}(r) -
      \widehat{\mu}_{T}\right)^{2}\right]. \label{eq:musigma}
\end{eqnarray}
Under the null it can be shown that $\widehat{\mu}_{T}$ and 
$\widehat{\sigma}_{T}^{2}$ consistently estimate $\mu_{T}$ and $\sigma_{T}^{2}$ if 
$M/T\rightarrow 0$ as $M\rightarrow \infty$ and
$T\rightarrow \infty$. This implies that under the null
\begin{eqnarray*}
 \frac{S_{T} - \widehat{\mu}_{T}}{\widehat{\sigma}_{T}} \Dcon \mathcal{N}(0,1) \qquad \textrm{
    as }M,T\rightarrow \infty.
\end{eqnarray*}
Of course, in practice $M$ is fixed, and $\var[S_{T} -
\widehat{\mu}_{T}]\approx (1+\frac{1}{2M})\sigma_{T}^{2}$. Thus 
a better finite sample approximation
uses (\ref{eq:sigmaT}) to give
\begin{eqnarray*}
\frac{S_{T} - \widehat{\mu}_{T}}{\sigma_{T}} \Dcon \mathcal{N}\left(0,\left[1+\frac{1}{2M}\right]\right).
\end{eqnarray*}
This together with $(2M-1)\widehat{\sigma}_{T}^{2}\Dcon
\chi^{2}_{2M-1}$, which is asymptotically independent of
$S_{T}-\mu_{T}$, gives that
 \begin{eqnarray}
\label{eq:ttt}
\frac{S_{T} - \widehat{\mu}_{T}}{\widehat{\sigma}_{T}} \Dcon \left(1+\frac{1}{2M}\right)^{1/2}t_{2M-1}.
\end{eqnarray}
However, for finite $T$, $S_{T}$ is positive and the finite sample distribution of $S_{T}$
 tends to be right skewed.  
To make $S_{T}$ more normal \cite{p:che-04} (see also
\cite{p:ter-03}), propose the power transform, $S_{T}^{\beta}$, where
$0 < \beta < 1$, which makes the distribution of $S_{T}^{\beta}$ more
normal than $S_{T}$. Since $S_{T}$ is asymptotically normal (see
(\ref{eq:AsymptoticNormal})) by using 
\cite{p:che-04} we have 
\begin{eqnarray*}
\frac{S_{T}^{\beta}  - \mu_{(\beta)}(\mu_{T},
\sigma_{T})}{\sigma_{(\beta)}(\sigma_{T})} \Dcon N(0,1)
\end{eqnarray*}
where 
\begin{eqnarray*}
\mu_{(\beta)}(\mu_{T},\sigma_{T})
 = \mu_{T}^{\beta} +
 \frac{1}{2}\beta(\beta-1)\mu_{T}^{\beta-2}\sigma_{T}^{2} \textrm{ and
 }
\sigma_{(\beta)}(\sigma_{T})= \beta\mu_{T}^{\beta-1}\sigma_{T}.
\end{eqnarray*}
Since $\mu_{T}$ and $\sigma_{T}$ are unknown we replace 
the above by
the estimators defined in (\ref{eq:musigma}). Using the same arguments
as those used in the derivation of (\ref{eq:ttt}) we have
 \begin{eqnarray}
\label{eq:ST2}
\frac{S_{T}^{\beta}  - \mu_{(\beta)}(\widehat{\mu}_{T},
\widehat{\sigma}_{T})}{\sigma_{(\beta)}(\widehat{\sigma}_{T})}\Dcon \left(1+\frac{1}{2M}\right)^{1/2}t_{2M-1}.
\end{eqnarray}
\cite{p:che-04} propose selecting $\beta$ to minimize the skewness of 
the statistic, so that the centralized third-order
moment is zero. Following the same calculations as those used in
\cite{p:che-04} this means using
\begin{eqnarray}
\label{eq:chen}
\beta = 1 - \frac{\mu_{T}\Ex[S_{T}-\mu_{T}]^{3}}{3\sigma_{T}^{4}}.
\end{eqnarray}
However, as the three terms (mean, variance and centralized third
moment) are unknown we estimate them using the orthogonal sample 
$\{S_{R,T}(r),S_{I,T}(r)\}_{r=1}^{M}$. 

We illustrate the above procedure with some simulations. 
Following \cite{p:det-pap-09}, we use the 
 linear bivariate time series 
\begin{eqnarray*}
X_{t} = 0.8X_{t-1} + \varepsilon_{t} \quad \textrm{ and }\quad Y_{t} = 0.8Y_{t-1} + \delta
Y_{t-2} +\eta_{t},
\end{eqnarray*}
where $\{(\varepsilon_{t},\eta_{t})\}_{t}$ are iid bivariate Gaussian random
variables with $\var[\varepsilon_{t}]=1$, $\var[\eta_{t}]=1$ and
 $\cov[\varepsilon_{t},\eta_{t}]=\rho$. If $\delta=0$
then the spectral densities of $\{X_{t}\}$ and $\{Y_{t}\}$ are the
same and the null hypothesis is true. For the alternative hypothesis
we use
$\delta = 0.1$ and $-0.1$.  We use the test statistic
$S_{T}$ defined in (\ref{eq:ST1}) and (\ref{eq:ST2}), where the
spectral density is estimated using the Daniell kernel. 
We use (i) $\beta = 0.25$ and (ii) an estimate of
(\ref{eq:chen}) (which we denote as $\widehat{\beta}$). 
In the simulations we use $T=128, 512$ and $1024$ over 500
replications. 

The results are reported in Table \ref{table:linear}, where all the tests are done
at the $5\%$ level.
From Table \ref{table:linear} we see that under the null the test
statistic seems to retain the 5\% level relatively well when $T=512$ and 
$T=1024$. There is, however, some over rejection when $T=128$.
We see that for $\rho = 0.1$ and $-0.1$ (the alternative is true),
that the test has power which grows as $T$ grows. We note that the 
estimated power transform $\widehat{\beta}$ tends, on average, to be larger than
$0.25$. Thus the distribution of $S_{T}^{\widehat{\beta}}$ tends to be
more right skewed than $S_{T}^{0.25}$. This may explain why the
proportion of rejection levels under the null using $S_{T}^{\widehat{\beta}}$ are 
a little larger than those with $S_{T}^{0.25}$. Comparing our
results to those reported in \cite{p:det-pap-09}, we see their frequency
bootstrap procedure performs a little better. There are two
possible explanations for this (i) they use a different test
statistic, based on ratios (ii) the power transform
may make the test a little conservative in rejecting the
null. However, it is interesting to note that a procedure with very
little computational expensive performs relatively well even against
bootstrap procedures. 

In the following section, we look again at the issue of testing. 
We have seen that by construction the orthogonal sample has approximately
the same variance as the statistic of interest. Furthermore, in many
testing procedures, under the null, both the test statistic and the
orthogonal sample have the same distribution. 
In this section, this property was exploited to estimate the mean and
variance of the test statistic. This in conjunction with the
t-distribution (with the appropriate number of degrees of
freedom) was used to obtain the p-value of the test statistic. 
An alternative approach is to use the
orthogonal sample to estimate the sampling distribution of the test
statistic under the null hypothesis. We investigate this in the
following section. 

\begin{table}
\centering
\scalebox{0.8}{
\begin{tabular}{|c|c|c|c|c|c|c|c|}
\hline
$\rho$ & $\delta$ & \multicolumn{2}{|c}{$T=128$ and $b=0.15$} &
\multicolumn{2}{|c}{$T=512$ and $b=0.1$} &
\multicolumn{2}{|c|}{$T=1024$ and $b=0.1$} \\
&   & $\widehat{\beta}$ & $\beta=0.25$ & $\widehat{\beta}$ & $\beta=0.25$ & $\widehat{\beta}$
   & $\beta=0.25$ \\
\hline
0.9 & 0.0 & 14.4 & 9.6 & 6.2 & 5.4 & 5.6 & 4.8 \\
      & 0.1 & 54.4 & 39.6 & 93.2 & 91.2 & 100 & 99.6 \\
      &-0.1 & 32.8 & 24.2 & 74.8 & 73.8 & 95.6 & 97.8 \\
\hline
0.5 & 0.0 & 13.6 & 8.4 & 5 & 4 & 3.2 & 3 \\
      & 0.1 & 32.8 & 27.2 &57.4 & 51.6 & 81.8 & 81.2 \\
      & -0.1 & 16.6 & 12.4 & 30.2 & 27.6 & 48 &  47 \\
\hline
0.0 & 0.0 & 12.8 & 8.6 & 4.4 & 3.8 & 4.8 & 4.4 \\
       & 0.1 & 26.8 & 20.4 &47.2 & 43 & 72.4 & 70 \\
      & -0.1 & 13.2 & 9.2 &19.4 & 17.7 & 33 & 31.6 \\
\hline 
-0.5 & 0.0 & 12.2 & 8 & 7.4 & 5.2 & 3.6 & 2.8 \\
       & 0.1 & 31.6 & 24.8 & 54 & 49.2 & 80 & 80.4 \\
       & -0.1 & 16.8 & 13.0 & 23.8 & 20.8 & 43.8 & 43.6 \\
\hline 
  -0.9 & 0.0 & 10.8 & 8 & 7.6 & 5.8 & 4.6 & 4.2 \\
          & 0.1 & 56.8 & 45.6 & 90 & 87.4 & 98.4 & 98.4 \\
          & -0.1 & 32.6 & 24.8 &  74.2 & 69.6 & 93.4 & 94 \\
\hline
\end{tabular}}
\caption{When $T=128$ we use $M=6$, when $T=512$ we use $M=12$ and
  when $T=1024$ we use $M=18$. \label{table:linear}}
\end{table}

%% file: 4_testing.tex
\section{Testing in Time Series}\label{sec:test}

Many test
statistics in time series can be formulated in terms of the parameters
$\{A_{}(\phi_{j})\}_{j}$ for some particular set of functions $\{\phi_{j}\}$,
 where under the null hypothesis
$A_{}(\phi_{j}) = 0$ for $j=1,\ldots,L$ and under the alternative
$A_{}(\phi_{j}) \neq 0$. This motivates the popular $\ell_{2}$
test statistic
\begin{eqnarray*}
S_{T} = T\sum_{j=1}^{L}|A_{T}(\phi_{j})|^{2}.
\end{eqnarray*}
In this section we use orthogonal samples to estimate the distribution of $S_{T}$ under the null
hypothesis.

By using the results in Section \ref{sec:quad} and Remark
\ref{remark:covariance}, equation (\ref{eq:sigmaj1j2}), we observe
that $\{A_{T}(\phi_{j})\}_{j=1}^{L}$ and for small $r$
$\{\sqrt{2}\Re A_{T}(\phi_{j};r)\}_{j=1}^{L}$ and $\{\sqrt{2}\Im A_{T}(\phi_{j};r)\}_{j=1}^{L}$
asymptotically have the same variance matrix.  In addition, under the null hypothesis that
$A_{}(\phi_{j}) = 0$ for $j=1,\ldots,L$, asymptotically 
$\{\sqrt{2}\Re
A_{T}(\phi_{j};r),\sqrt{2}\Im A_{T}(\phi_{j};r)\}$ and
$A_{T}(\phi_{j})$ have the same mean and limiting Gaussian
distribution. This suggests that the distribution of $S_{T}$ under the
null can be approximated by the empirical distribution of the
corresponding orthogonal sample. 
 Based on the above observations we define the 
orthogonal sample associated with $S_{T}$ as 
\begin{eqnarray*}
S_{T,R}(r) = 2T\sum_{j=1}^{L}|\Re A_{T}(\phi_{j};r)|^{2} \textrm{
  and }S_{T,I}(r) = 2T\sum_{j=1}^{L}|\Im A_{T}(\phi_{j};r)|^{2}
\textrm{ for } 1\leq r \leq M.
\end{eqnarray*}
In the theorem below we show that under the null
hypothesis $H_{0}:A_{}(\phi_{j}) = 0$ for $1\leq j \leq L$, the 
asymptotic sampling properties of 
$S_{T}$, $S_{T,R}(r)$ and $S_{T,I}(r)$ are equivalent. 

\begin{theorem}\label{theorem:orthogonalQ}
Suppose Assumption \ref{assum:p} holds with $p=16$. Furthermore, we assume $\{\phi_{j}\}$ are
Lipschitz continuous functions and $\Re A_{T}(\phi_{j}) =
A_{T}(\phi_{j})$. 
Let $V_{j_{1},j_{2}}$ be defined as
 \begin{eqnarray*}
V_{j_{1},j_{2}} &=& \frac{1}{2\pi}\int_{0}^{2\pi}f(\omega)^{2}
\left[\phi_{j_{1}}(\omega)\overline{\phi_{j_{2}}(\omega)} +\phi_{j_{1}}(\omega)\overline{\phi_{j_{2}}(-\omega)}\right]
d\omega + \nonumber\\
&&\frac{1}{2\pi}\int_{0}^{2\pi}\int_{0}^{2\pi}\phi_{j_{1}}(\omega_{1})\overline{\phi_{j_{2}}(\omega_{2})}
f_{4}(\omega_{1},-\omega_{1},-\omega_{2})d\omega_{1}d\omega_{2}+ O(T^{-1}).
\end{eqnarray*}
Then we have 
\begin{itemize}
\item[(i)] \underline{The mean}
\begin{itemize}
\item[(a)] Under the null hypothesis that $A_{}(\phi_{j}) = 0$
  for $1\leq j \leq L$  we have 
\begin{eqnarray*}
\Ex[S_{T}] = \sum_{j=1}^{L}V_{j,j} + O(T^{-1}). 
\end{eqnarray*}
However, if
for at least one $1\leq j\leq L$ $A_{}(\phi_{j}) \neq 0$, then $\Ex[Q_{T}] = O(T)$.
\item[(b)] Under both the null and alternative and for $0<r<T/2$ we have
\begin{eqnarray*}
\Ex[S_{T,R}(r)] = \sum_{j=1}^{L}V_{j,j} + O(|r|T^{-1}) \textrm{ and }
\Ex[S_{T,I}(r)] = \sum_{j=1}^{L}V_{j,j} + O(|r|T^{-1}).
\end{eqnarray*}
\end{itemize}
\item[(ii)] \underline{The covariance}
\begin{itemize}
\item[(a)] Under the null hypothesis, 
$\var[S_{T}] = 2\sum_{j_{1},j_{2}=1}^{L}V_{j_{1},j_{2}}^{2} + O(T^{-1})$ 
\item[(b)] Under both the null and alternative hypothesis where $1\leq
  r_{1},r_{2}< T/2$ we have  
\begin{eqnarray*}
cov[S_{T,R}(r_{1}),S_{T,R}(r_{2})] &=& 
\left\{
\begin{array}{cc}
2\sum_{j_{1},j_{2}=1}^{L}V_{j_{1},j_{2}}^{2} + O(|r|T^{-1})  & r_{1} =
r_{2} (= r)\neq 0 \\
O(T^{-1})
\end{array}
\right.\\
cov[S_{T,I}(r_{1}),S_{T,I}(r_{2})] &=& 
\left\{
\begin{array}{cc}
2\sum_{j_{1},j_{2}=1}^{L}V_{j_{1},j_{2}}^{2} + O(|r|T^{-1})  & r_{1} =
r_{2} (= r)\neq 0 \\
O(T^{-1})
\end{array}
\right. \\
cov[S_{T,R}(r_{1}),S_{T,I}(r_{2})] &=& O(T^{-1}).
\end{eqnarray*}
\end{itemize}
\item[(iii)] \underline{Higher order cumulants} Suppose 
 Assumption \ref{assum:p} holds with the order $2p$.
Let $\cum_{p}$ denote the $p$th order cumulant of a random variable. Then
under the null hypothesis 
\begin{eqnarray*}
|\cum_{p}(S_{T}) - \cum_{p}(S_{T,R}(r))| = O(|r|T^{-1}), \quad |\cum_{p}(S_{T}) - \cum_{p}(S_{T,I}(r))| = O(|r|T^{-1}).
\end{eqnarray*}
\end{itemize}
\end{theorem}
We observe that the above theorem implies under the null, $S_{T}$, $S_{T,R}(r)$ and
$S_{R,I}(r)$ asymptotically have equivalent mean, variance and higher
order cumulants. Furthermore, under 
the alternative the asymptotic mean and variance of $S_{T,R}$ and
$S_{T,I}$ are finite and bounded with $M\rightarrow \infty$ as
$T\rightarrow \infty$. 
Therefore motivated by these results we define the empirical distribution
\begin{eqnarray}
\label{eq:QT-def-dist}
\widehat{F}_{M,T}(x) =
\frac{1}{2M}\left(\sum_{r=1}^{M}\left[I(S_{T,R}(r)\leq x) +
    I(S_{T,I}(r)\leq x)\right]\right). 
\end{eqnarray}
To do the test we use $\widehat{F}_{M,T}(x)$ as an approximation of the
distribution of $S_{T}$ under the null hypothesis. 
We reject the null at the $\alpha \%$-level if
$1-\widehat{F}_{M,T}(S_{T})< \alpha\%$. We note that under the
alternative that at least one $j=1,\ldots,L$
$A(\phi_{j})\neq 0$, then 
$S_{T} = O_{p}(T)$. By Theorem \ref{theorem:orthogonalQ}(ii) the variance of 
 $S_{T,R}(r)$ and $S_{T,I}(r)$ is finite and uniformly bounded for all
 $r$ and $T$. This implies that 
 $1-\widehat{F}_{M,T}(S_{T})\Pcon 0$ as  $M$ and $T\rightarrow \infty$, thus
 giving the procedure power. 
To show a Glivenko-Cantelli type result of the form $\sup_{x\in \mathbb{R}}|\widehat{F}_{M,T}(x)-F(x)|\AScon
0$ as $M\rightarrow \infty$ and $T\rightarrow \infty$, where $F$ denotes the limiting distribution of $Q_{T}$ under the
null hypothesis ($F$ is a generalised chi-squared) is beyond the scope of the current paper. 
Based on the simulations in Section \ref{sec:sim}, we conjecture that
this result is true. 

We now apply this procedure to test for uncorrelatedness and goodness
of fit. 

\subsection{A Portmanteau test for uncorrelatedness}\label{sec:port}

Let us suppose we observe the stationary time series $\{X_{t}\}$. 
The classical test for serial correlation assumes that under the null hypothesis
the observations are independent, identically distributed (iid) random
variables. In this case
the classical Box-Pierce statistic is defined as 
\begin{eqnarray}
\label{eq:BP}
\widetilde{Q}_{T} = \frac{T}{\widetilde{c}_{T}(0)^{2}}\sum_{j=1}^{L}\left|\widetilde{c}_{T}(j)\right|^{2},
\end{eqnarray}
where $\widetilde{c}_{T}(j)$ is defined in Example \ref{example:1}(a).
If the null holds, then $\widetilde{Q}_{T}$ is asymptotically a chi-square distribution with $L$ degrees of
freedom. However, if the intention is to test for uncorrelatedness
without the additional constraint of independence, then it can be shown that 
\begin{eqnarray}
\label{eq:cov-cov}
T\cov[\widetilde{c}_{T}(j_{1}),\widetilde{c}_{n}(j_{2})]  
 &=& c(0)^{2}\delta_{j_{1},j_{2}} +
 c(0)^{2}\delta_{j_{1},j_{2}}\delta_{j_{1},0}  +  \sum_{k=-\infty}^{\infty}\kappa_{4}(j_{1},k,k+j_{2}),
\end{eqnarray}
where $\delta_{j_{1},j_{2}}$ is the dirac-delta function (see
\cite{b:bro-dav-91}, Chapter 7, for the derivation in the case of a
linear time series and  \cite{p:rom-tho-96} in the general case). 
Consequently, under the null of uncorrelatedness, the distribution of $\widetilde{Q}_{T}$
 is not a standard chi-square.  

\cite{p:die-86}, \cite{p:wei-86}, \cite{p:rob-91}, \cite{p:ber-93} and
\cite{p:esc-lob-09} avoid some of these issues by placing stronger
conditions on the time series and assume that under the
null hypothesis the time series are martingales differences (thus
uncorrelated). This implies that the
fourth order cumulant term in (\ref{eq:cov-cov}) is zero in the case
that $j_{1}\neq j_{2}$, which induces asymptotic uncorrelatedness
between the sample covariances. Based on this observation they propose
the robust Portmanteau test 
\begin{eqnarray}
\label{eq:robustP}
Q_{T}^{*} = T\sum_{j=1}^{L}\frac{\left|\widetilde{c}_{T}(j)\right|^{2}}{\widehat{\tau}_{j}},
\end{eqnarray}
where
$\widehat{\tau}_{j}=\frac{1}{n-j}\sum_{t=j+1}^{n}(X_{t}-\bar{X})^{2}(X_{t-j}-\bar{X})^{2}$. 
Under the null of martingale differences $Q_{T}^{*}$ is asymptotically
has $\chi^{2}$-distributed 
with $L$-degrees of freedom. 
However, if the intention is to test for  uncorrelatedness, {\it without}
additional assumptions on the structure, then, even under the null,
$Q_{T}^{*}$ will have a generalised chi-squared distribution, whose
parameters are difficult to estimate. This problem motivated
\cite{p:rom-tho-96} (using the block bootstrap) and 
\cite{p:lob-01} (who developed the
method of self-normalisation) to test for uncorrelatedness under
these weaker conditions.

We will use orthogonal samples to estimate the distribution of
Portmanteau statistic under the null that $H_{0}:c(j)=0$ for all
$1\leq j\leq L$. 
We recall from Example \ref{example:1}(a)
that $A_{T}(e^{ij\cdot})$ is an estimator of the autocovariance
$\widehat{c}_{T}(j)$. Therefore, to 
test for uncorrelatedness at lag $j=1,\ldots,L$ we define the test statistic
\begin{eqnarray}
\label{eq:QT-def}
Q_{T} = T\sum_{j=1}^{L}|A_{T}(e^{ij\cdot})|^{2}.
\end{eqnarray}
Using 
$\{\sqrt{2}\Re A_{T}(e^{ij\cdot};r), \sqrt{2}\Im A_{T}(e^{ij\cdot};r)
;r=1,\ldots,M\}$ we define the orthogonal sample associated with 
$Q_{T}$ as 
\begin{eqnarray*}
Q_{T,R}(r) = 2T\sum_{j=1}^{L}|\Re A_{T}(e^{ij\cdot};r)|^{2} \textrm{
  and }Q_{T,I}(r) = 2T\sum_{j=1}^{L}|\Im A_{T}(e^{ij\cdot};r)|^{2}
\textrm{ for } 1\leq r \leq M.
\end{eqnarray*}
The above orthogonal sample is used to define the empirical
distribution, like that defined in (\ref{eq:QT-def-dist}), we denote
this as $\widehat{F}_{Q,M,T}(x)$.  
We reject the null at the $\alpha \%$-level if
$1-\widehat{F}_{Q,M,T}(Q_{T})< \alpha\%$. Results of the corresponding
simulation study is given in Section \ref{sec:sim}, where we apply the proposed methodology to a 
wide class of uncorrelated processes. 

\subsection{Testing for goodness of fit}\label{sec:good}

In this section we describe how orthogonal samples can be applied 
to testing for goodness of fit. Given that $f$ is the spectral density
of the observed time series, our objective  is to test $H_{0}:f(\omega) =
g(\omega;\theta)$ for all $\omega\in [0,2\pi]$ against
$H_{A}:f(\omega)\neq g(\omega;\theta)$ for some $\omega\in [0,2\pi]$. 
Typically this is done by fitting the model to the data and applying
the Portmanteau test to the
residuals -- in either the time or frequency domain (cf. \cite{p:mil-81},
\cite{p:hon-96}).
In Example \ref{example:1}(c) it  was observed that
$A_{T}(e^{ij\cdot}g(\cdot;\theta)^{-1})$ is an estimator of the
covariance of the residuals at lag $j$.  
Under the null hypothesis that $g$ is the true spectral density,
then the residual covariance $A_{T}(e^{ij\cdot}g(\cdot;\theta)^{-1})$ is estimating
zero. Using this observation
\cite{p:mil-81}  defines the  statistic  
\begin{eqnarray}
\label{eq:GT-def}
G_{T} = T\sum_{j=1}^{L}\left|A_{T}(e^{ij\cdot}g(\cdot;\theta)^{-1})\right|^{2}
\end{eqnarray}
to test for goodness of fit. 
Using $G_{T}$ we define the orthogonal sample 
\begin{eqnarray*}
G_{T,R}(r) = 2T\sum_{j=1}^{L}|\Re A_{T}(e^{ij\cdot}g(\cdot;\theta)^{-1};r)|^{2} \textrm{
  and }G_{T,I}(r) = 2T\sum_{j=1}^{L}|\Im A_{T}(e^{ij\cdot}g(\cdot;\theta)^{-1};r)|^{2}.
\end{eqnarray*}
Using Theorem \ref{theorem:orthogonalQ}, under the null hypothesis,
$G_{T}$, $G_{T,R}(r)$ and $G_{T,I}(r)$ asymptotically share the same
sampling properties when $r$ is small. 
We use (\ref{eq:QT-def-dist}) to define the corresponding empirical
distribution, which we denote as $\widehat{F}_{G,M,T}(x)$.
We reject the null, that the spectral density is $g(\cdot)$ at the $\alpha \%$-level if
$1-\widehat{F}_{G,M,T}(G_{T})< \alpha\%$.

%% file: 4_a_simulations.tex

\subsection{Simulations}\label{sec:sim}

In the following section we assess the tests described above through
some simulations.
All  tests are done at the $\alpha=5\%$ and $\alpha=10\%$ nominal
levels. The methods are compared to the block bootstrap method,  where
to obtain the bootstrap critical values $1000$ bootstrap samples were taken.
Throughout this section we let $\{Z_{t}\}$ and $\{\varepsilon_{t}\}$
 denote independent, identically distributed standard normal random
 variables and
 chi-square with one degree of freedom random variables respectively.

\subsection{Example 3: Testing for uncorrelatedness}\label{sec:sim-cor}

In this section we illustrate the test for uncorrelatedness using the
orthogonal sample method described in
Section \ref{sec:port}. We use the test statistic $Q_{T}$ (defined in
(\ref{eq:QT-def})), using $L=5$,  
and obtain the critical values using the
empirical distribution function, $\widehat{F}_{Q,M,T}$ defined in (\ref{eq:QT-def-dist}).
We compare our method with the regular Box-Pierce statistic (defined
in (\ref{eq:BP})) and the 
robust Portmanteau test statistic defined in
(\ref{eq:robustP}), for both these methods we obtained the critical
values using the $\chi^{2}$ distribution with five degrees of
freedom. In addition, we compare our method to the results
of the bootstrap test where the critical values are obtained using the
block bootstrap procedure. Namely,  the critical values for $Q_{T}$
are obtained using the centralised empirical distribution constructed
with samples from the block bootstrap procedure, with block bootstrap length $B=5,10$
and $20$. 

To select $M$ in the orthogonal sample method we use the average
squared criterion described in Section \ref{sec:cross}. More precisely, we
focus on the sample covariance at lag one and 
choose $M = \arg\min_{M\in \mathcal{S}}C_{\phi}(M)$, where 
\begin{eqnarray}
\label{eq:mathcalCM}
\mathcal{C}_{\phi}(M) = \frac{4}{T}\sum_{r=1}^{T/4}\left(\frac{\left|\sqrt{T}A_{T}(e^{i\cdot};r)\right|^{2}}{
\widehat{V}_{M}(\omega_{r})}-1 \right)^{2}\textrm{ with }
\widehat{V}_{M}(\omega_{r}) = \frac{T}{M}\sum_{s=1+r}^{M+r}|A_{T}(e^{i\cdot};s)|^{2}.
\end{eqnarray}
and $\mathcal{S} = \{10,\ldots,30\}$.

\subsubsection*{Models under the null of no correlation}

The first two models we consider are iid random variables which follow
a standard normal distribution and a t-distribution with five degrees
of freedom. The third model is the two-dependent model $X_{3,t} =
Z_{t}Z_{t-1}$.
The fourth model we consider is the non-linear, non-martingale, 
uncorrelated model,  defined in \cite{p:lob-01}, where
$X_{4,t} = Z_{t-1}Z_{t-2}\left(Z_{t-1}+Z_{t}+1\right)$.
The fifth model we consider is the ARCH$(1)$ process
$X_{5,t}$, where
\begin{eqnarray*}
X_{5,t} = \sigma_{5,t}Z_{t} \qquad \sigma_{5,t}^{2} = 1+0.8X_{5,t-1}^{2}.
\end{eqnarray*}
The sixth model is $X_{6,t} = |X_{5,t}|V_{t}$ where $\{X_{5,t}\}$
 and $\{V_{t}\}$ are independent of each other, $X_{5,t}$ is the ARCH
 process described above and $V_{t}$ is an uncorrelated non-causal time
 series defined by
\begin{eqnarray*}
V_{t} = \sum_{j=0}^{\infty}a^{j}\varepsilon_{t-j} - \frac{a}{1-a^{2}}\varepsilon_{t+1},
\end{eqnarray*}
where $a=0.8$.  The seventh model we consider is a `pseudo-linear'
non-causal, uncorrelated time series with ARCH innovations defined by 
\begin{eqnarray*}
X_{7,t} = \sum_{j=0}^{\infty}b_{1}^{j}U_{1,t-j} -
\frac{b_{1}}{1-b_{1}^{2}}U_{1,t+1}, \quad 
U_{1,t} = \sum_{j=0}^{\infty}b_{2}^{j}U_{2,t-j} - \frac{b_{2}}{1-b_{2}^{2}}U_{2,t+1}
\end{eqnarray*}
where $U_{2,t} = \sigma_{2,t}Z_{t}$ with $\sigma_{t} =
1+0.5U_{2,t-1}^{2}$, $b_{1}=-0.8$ and $b_{2}=-0.6$. 
Finally, the eighth model we consider is the periodically stationary model
defined in \cite{p:pol-97}, 
$X_{7,t} = s_{t}X_{3,t}$ and $s_{t}$ is a deterministic sequence of
period $12$,  where the elements are $\{1,1,1,2,3,1,1,1,1,2,4,6\}$
(this time series does not satisfy our stationary assumptions).
We used the sample sizes $T=100$ and $T=500$.

The results are given in Tables \ref{table:1} and \ref{table:2}.
We observe that overall the orthogonal sampling method keeps to the nominal
level, with a mild inflation of the type I error for independent data
(normal and t-distribution). It is likely that this is because the $5$th and $10$th
quantile is estimated using a maximum of 60 points,
since the order selection set  is $\mathcal{S} = \{10,\ldots,30\}$
(often it is a lot less than $60$). It is a little surprising that the type I errors
for model $\{X_{t,4}\}$ lie far below the nominal level.  
As expected, the regular Box-Pierce
statistic keeps the nominal level well for the iid data, but cannot
control the type I error when the data is uncorrelated but not iid. Suprisingly, the robust Portmanteau test is able to keep
the type I error in most cases, the exception being the pseudo-linear
model, $X_{7,t}$, where there is a mild inflation of the type I error.
In the case of the Block Bootstrap for $T=100$ the performance depends
on the size of the block. For $B=5$ and $B=10$ the type I error is
below the nominal level, whereas for $B=20$ the type I error tends to
be around and above the nominal level. However, when $T=500$ the block bootstrap
is consistently below the nominal level for $B=5,10$ and $20$. This
suggests a larger block length should be used. It is quite possible
that more accurate critical values, which are less sensitive to block
length, can be obtained using the fixed-$b$ bootstrap. However,
the fixed-$b$ bootstrap was not included in the study as  the
aim in this section is to compare different procedures which are
simple and fast to implement using routines that already exist in R. 

\begin{table}[h!]
\centering
\scalebox{0.8}{
\begin{tabular}{|c||c|c||c|c|c|c|c|c|c|c|c|c|}
\hline
 Model & \multicolumn{2}{c||}{Orthogonal $Q_{T}$} & \multicolumn{2}{c|}{Regular $\widetilde{Q}_{T}$} & \multicolumn{2}{c|}{Robust $Q_{T}^{*}$} &
 \multicolumn{6}{c|}{Block Bootstrap} \\
 & & & & & & & \multicolumn{2}{c|}{B=5} & \multicolumn{2}{c|}{B=10} &\multicolumn{2}{c|}{B=20} \\
\hline \hline           
       & 5\% & 10\% & 5\% & 10\% & 5\% & 10\% & 5\% & 10\% & 5\% &
       10\% & 5\% & 10\% \\
\hline \hline
Normal & 6.52 & 11.1 & 4.1 & 8.34 & 5.42 & 10.18 & 0.0 & 0.1 &
1.14 & 4.36 & 5.14 & 11.94 \\
$t_{5}$  & 6.34 & 11.42 & 4.08 & 8.46 & 4.92 & 10.5 & 0.0 & 0.08 &
0.96 & 4.12 & 4.74 & 10.70 \\
$X_{3,t}=Z_{t}Z_{t-1}$ & 5.02 & 9.44 & 10.66 & 16.64 & 4.38 & 9.52 & 0.14 &
0.66 & 1.2 & 4.68 & 4.14 & 11.14 \\
$X_{4,t}$ & 0.86 & 1.82 & 3.5 & 4.86 & 1.1 & 2.32 & 0.06 & 0.3 & 0.22 &
1.0 & 0.64 & 2.26 \\
$X_{5,t}$ & 4.26 & 8.14 & 23.56 & 31.6 &  5.4 & 9.98 & 0.14 & 1.06 & 1.22 &
4.96 & 3.54 & 11.00\\
$X_{6,t}$ & 3.16 & 6.42 & 17.64 & 24.22 & 4.20 & 8.38 & 0.08 & 0.4 & 0.6
& 2.92 & 2.08 & 7.78 \\
$X_{7.t}$ & 5.1 & 10.46 & 13.22 & 20.46 & 6.88 & 12.8 & 0.16 & 0.86 &
1.18 & 5 & 4.56 & 11.88 \\
$X_{8,t}$ & 4.46 & 8.36 & 8.2 & 13.18 & 3.5 & 8.38 & 0.04 & 0.58 & 0.74 &
4 & 3.1 & 10.48 \\
\hline
\end{tabular}}
\caption{Test for uncorrelatedness,  under the null hypothesis, $T=100$ over 5000
  replications. \label{table:1}}
\end{table}

\begin{table}[h!]
\centering
\scalebox{0.8}{
\begin{tabular}{|c||c|c||c|c|c|c|c|c|c|c|c|c|}
\hline
 Model & \multicolumn{2}{c||}{Orthogonal $Q_{T}$} & \multicolumn{2}{c|}{Regular $\widetilde{Q}_{T}$} & \multicolumn{2}{c|}{Robust $Q_{T}^{*}$} &
 \multicolumn{6}{c|}{Block Bootstrap} \\
 & & & & & & & \multicolumn{2}{c|}{B=5} & \multicolumn{2}{c|}{B=10} &\multicolumn{2}{c|}{B=20} \\
\hline \hline           
       & 5\% & 10\% & 5\% & 10\% & 5\% & 10\% & 5\% & 10\% & 5\% &
       10\% & 5\% & 10\% \\
\hline \hline
Normal & 5.9 & 11.1 & 4.56 & 9.44 & 4.74 & 9.86 & 0.08 & 0.44 & 1.48 &
4.22 & 3.7 & 8.78 \\
$t_{5}$  & 6.1 & 10.82 & 4.8 & 9.58 & 4.92 & 9.98 & 0.04 & 0.34 & 1.24
& 4.18 & 3.3 & 8.66 \\
$X_{3,t}=Z_{t}Z_{t-1}$ & 5.00 & 9.82 & 15.26 & 22.26 & 4.9 & 9.16 & 0.7 &
2.52 & 2.5 & 6.56 & 3.6 & 9.16 \\
$X_{4,t}$ & 1.0 & 1.86 & 30.52 & 38.56 & 5.6 & 11.16 & 1.56 & 4.48 &
2.72 & 7.94 & 3.1 & 10.02 \\
$X_{5,t}$ & 3.76 & 7.06 & 49.86 & 58.76 & 4.82 & 9.34 & 1.06 & 4.10 & 2.08
& 6.9 & 2.52 & 7.96 \\
$X_{6,t}$ & 2.88 & 6.22 & 42.48 & 50.46 & 3.64 & 7.24 & 0.64 & 2.4 & 1.24
& 4.68 & 1.5 & 6.62 \\
$X_{7,t}$ & 4.48 & 8.88 & 20.38 & 28.32 & 6.02 & 11.46 & 0.78 & 2.58 &
1.52 & 5.04 & 2.78 & 7.5 \\
$X_{8,t}$ & 5.28 & 9.46 & 15.08 & 20.46 & 4.12 & 8.54 & 1.28 & 3.82 &
3.04 & 7.36 & 4.34 & 9.92 \\
\hline
\end{tabular}}
\caption{Test for uncorrelatedness,  under the null hypothesis, $T=500$ over 5000
  replications. \label{table:2}}
\end{table}

\subsubsection*{Models under the alternative of correlation}

To access the empirical power of the test we consider three different
models. The first model is the Gaussian autoregressive process $Y_{1,t}$, where 
$Y_{1,t} = -0.2Y_{1,t-1} +Z_{t}$.
The second model is $Y_{2,t} = Y_{1,t}|U_{2,t}|$, where $\{Y_{1,t}\}$ and $\{U_{t,2}\}$
are independent of each other, $\{Y_{1,t}\}$ is defined above and 
$\{U_{t,2}\}$ is the ARCH model defined in the
previous section. Finally, the third model is $Y_{3,t} = U_{3,t}|U_{2,t}|$, where $\{U_{2,t}\}$ and $\{U_{3,t}\}$
are independent of each other, $\{U_{t,2}\}$ is the ARCH model defined in the
previous section and $\{U_{3,t}\}$ is the Gaussian autoregressive process 
$U_{3,t} = 0.5U_{4,t-1}+Z_{t}$. We used the sample sizes $T=100$, $T=200$ and $T=500$.

The result are given in Tables  \ref{table:3}, \ref{table:4} and \ref{table:5}.
The power for most of the methods are relatively close. Though it is
not surprising that the regular
Box-Pierce statistic has the largest power, since it also has the
largest inflated type I errors.  Overall, in terms of
power, the orthogonal sampling test and the robust Portmanteau test tend
to have more power than the Block Bootstrap test, especially when the
sample size is small.

\begin{table}[h!]
\centering
\scalebox{0.8}{
\begin{tabular}{|c||c|c||c|c|c|c|c|c|c|c|c|c|}
\hline
 Model & \multicolumn{2}{c||}{Orthogonal $Q_{T}$} & \multicolumn{2}{c|}{Regular $\widetilde{Q}_{T}$} & \multicolumn{2}{c|}{Robust $Q_{T}^{*}$} &
 \multicolumn{6}{c|}{Block Bootstrap} \\
 & & & & & & & \multicolumn{2}{c|}{B=5} & \multicolumn{2}{c|}{B=10} &\multicolumn{2}{c|}{B=20} \\
\hline \hline           
       & 5\% & 10\% & 5\% & 10\% & 5\% & 10\% & 5\% & 10\% & 5\% &
       10\% & 5\% & 10\% \\
\hline \hline
$Y_{1,t}$ & 27.06 & 38.88 & 29.52 & 40.84 & 28.60 & 40.50 & 2.62 & 9.4 &
14.12 & 29.34 & 25.74 & 43.22 \\  
$Y_{2,t}$ & 12.68 & 20.58 & 21.78 & 30.6 & 11.08 & 18.68 & 0.5 & 3.3
& 3.94 & 13.84 & 9.96 & 24.02 \\
$Y_{3,t}$ & 55.7 & 68.6 & 71.94 & 79.58  & 61.8 & 72.32 & 18.36 & 41.46
& 41.62 & 65.14 & 52.86 & 74.30 \\
\hline
\end{tabular}}
\caption{Test for uncorrelatedness, under the alternative hypothesis, $T=100$ over 5000
  replications.\label{table:3}}
\end{table}

\begin{table}\centering
\scalebox{0.8}{
\begin{tabular}{|c||c|c||c|c|c|c|c|c|c|c|c|c|}
\hline
 Model & \multicolumn{2}{c||}{Orthogonal $Q_{T}$} & \multicolumn{2}{c|}{Regular $\widetilde{Q}_{T}$} & \multicolumn{2}{c|}{Robust $Q_{T}^{*}$} &
 \multicolumn{6}{c|}{Block Bootstrap} \\
 & & & & & & & \multicolumn{2}{c|}{B=5} & \multicolumn{2}{c|}{B=10} &\multicolumn{2}{c|}{B=20} \\
\hline \hline           
       & 5\% & 10\% & 5\% & 10\% & 5\% & 10\% & 5\% & 10\% & 5\% &
       10\% & 5\% & 10\% \\
\hline \hline
$Y_{1,t}$ & 54.70 & 67.64 & 58& 69.06 & 55.78 & 67.18 & 18.58 & 36.16 &
41.10 & 59.84 & 48.64 & 66.54 \\
$Y_{2,t}$ & 21.98 & 32.4 & 37.44 & 47.18 & 18.04 & 27.68 & 4.14 &
13.32 & 11.34 & 27.04 & 16.30 & 33.74 \\
$Y_{t,3}$ & 87.04 & 92.64 & 95.72 & 97.28 & 85.84 & 91.22 & 74.5 &
87.96 & 84.42 & 93.74 & 84.54 & 94.58 \\
\hline 
\end{tabular}}
\caption{Test for uncorrelatedness under the alternative hypothesis, $T=200$ over 5000
  replications. \label{table:4}}
\end{table}

\begin{table}[h!]
\centering
\scalebox{0.8}{
\begin{tabular}{|c||c|c||c|c|c|c|c|c|c|c|c|c|}
\hline
 Model & \multicolumn{2}{c||}{Orthogonal $Q_{T}$} & \multicolumn{2}{c|}{Regular $\widetilde{Q}_{T}$} & \multicolumn{2}{c|}{Robust $Q_{T}^{*}$} &
 \multicolumn{6}{c|}{Block Bootstrap} \\
 & & & & & & & \multicolumn{2}{c|}{B=5} & \multicolumn{2}{c|}{B=10} &\multicolumn{2}{c|}{B=20} \\
\hline \hline           
       & 5\% & 10\% & 5\% & 10\% & 5\% & 10\% & 5\% & 10\% & 5\% &
       10\% & 5\% & 10\% \\
\hline \hline
$Y_{1,2}$ & 94.86 & 97.44 & 95.94 & 97.84 & 95.24 & 97.60 & 82.30 & 91.74
& 92.96 & 96.48 & 93.76 & 97.28 \\ 
$Y_{2,t}$ & 49.50 & 60.58 & 69.60 & 77.24 & 37 & 49.90 & 29.92 &
45.30 & 41.48 & 58.28 & 42.2 & 60.98 \\
$Y_{3,t}$ & 98.86 & 99.36 & 99.94 & 99.98 & 98.5 & 99.16 & 99.2 & 99.68
& 98.98 & 99.82 & 97.94 & 99.82 \\
\hline
\end{tabular}}
\caption{Test for uncorrelatedness under the alternative hypothesis, $T=500$ over 5000
  replications.\label{table:5}}
\end{table}

\subsection{Example 4: Goodness of fit test}\label{sec:sim-good}

In this section we illustrate the goodness of fit test using the method described in
Section \ref{sec:good} to test 
 $H_{0}:f(\omega) =
g(\omega;\theta)$ for all $\omega\in [0,2\pi]$ against
$H_{A}:f(\omega)\neq g(\omega;\theta)$ for some $\omega\in
[0,2\pi]$. We use the test statistic 
$G_{T}$ (defined in
(\ref{eq:GT-def})), with $L=5$.
We obtain the critical values using the
estimated distribution function, $\widehat{F}_{G,M,T}$ defined in  (\ref{eq:QT-def-dist}).
We compare our method to the results
of the bootstrap test where the critical values are obtained using the
block bootstrap procedure. We use the block bootstrap length $B=5,10,20,30$
and $40$. 

To select $M$ in the orthogonal sample method, we focus on $j=1$ and use 
$M = \arg\min_{M\in \mathcal{S}}C_{\phi}(M)$, where $C_{\phi}(M)$ is
defined similar to (\ref{eq:mathcalCM}) and $\mathcal{S} = \{10,\ldots,30\}$.

\subsubsection*{Models under the null hypothesis}

The first model is the Gaussian autoregressive process
\begin{eqnarray*}
X_{0.6,t}^{G} = 0.6X_{0.6,t-1}^{G}+Z_{t},
\end{eqnarray*}
where $\{Z_{t}\}$ are iid standard normal random variables,  
with spectral density $f(\omega)=g(\omega;\theta) =
(2\pi)^{-1}|1-0.6e^{i\omega}|^{2}$. The second model is the
non-Gaussian autoregressive process
\begin{eqnarray*}
X_{0.6,t}^{\chi}  = 0.6X_{0.6,t-1}^{\chi}+\varepsilon_{t},
\end{eqnarray*}
where $\{\varepsilon_{t}\}_{t}$ are chi-square distributed random
variables with one degree of freedom.
The spectral density is $f(\omega)=g(\omega;\theta) =
2(2\pi)^{-1}|1-0.6e^{i\omega}|^{2}$. The third model is the non-Gaussian
autoregressive process 
\begin{eqnarray*}
X_{0.9,t}^{\chi}  = 0.9X_{0.9,t-1}^{\chi}+\varepsilon_{t},
\end{eqnarray*}
where $\{\varepsilon_{t}\}$ is defined above with spectral density $f(\omega)=g(\omega;\theta) =
2(2\pi)^{-1}|1-0.9e^{i\omega}|^{2}$.
We used the sample sizes $T=100$ and $T=500$.

The result are given in Tables  \ref{table:6} and \ref{table:7}.
We observe that the orthogonal sampling method keeps to the nominal
level for both $T=100$ and $T=500$, with an underestimation of the
type I error for model $\{X_{0,9,t}^{\chi}\}$. On the other hand for $T=100$, the block
bootstrap underestimates the nominal level when the block is too small
($B=5$ and $10$) and over inflates the type I error when the block is
too large ($B=40$ and sometimes $B=30$). The ideal block length seems to be somewhere
between $B=20$ to $40$. In the
case that $T=500$, the nominal level is underestimated for all the
block lengths considered.

\begin{table}[h!]
\centering
\scalebox{0.8}{
\begin{tabular}{|c||c|c||c|c|c|c|c|c|c|c|c|c|}
\hline
 Model & \multicolumn{2}{c||}{Orthogonal $Q_{T}$} & 
 \multicolumn{10}{c|}{Block Bootstrap} \\
 & & & \multicolumn{2}{c|}{B=5} & \multicolumn{2}{c|}{B=10}
 &\multicolumn{2}{c|}{B=20} &\multicolumn{2}{c|}{B=30}
 &\multicolumn{2}{c|}{B=40}\\
\hline \hline           
       & 5\% & 10\% & 5\% & 10\% & 5\% & 10\% & 5\% & 10\% & 5\% & 10\%& 5\% & 10\%\\
\hline 
$X_{0.6,t}^{G}$ & 2.32 & 4.94 & 0.00 & 0.00 & 0.18 & 0.96 & 2.18 &  7.06 &4.66
& 11.24 & 8.64 & 17.84\\
$X_{0.6,t}^{\chi}$ & 2.24 & 4.4 & 0 & 0.02 & 0.04 & 0.26 & 0.94 & 3.48 & 2.24 &
6.82 & 4.94 & 13.00\\
$X_{0.9,t}^{\chi}$ & 0.96 & 1.74 & 0 & 0 & 0 & 0.02 & 0.04 & 0.26 &  0.2 & 0.96 & 0.56 & 2.64\\
\hline 
\end{tabular}}
\caption{Goodness of fit test, under the null, T=100 over 5000 replications\label{table:6}}
\end{table}

\begin{table}[h!]
\centering
\scalebox{0.8}{
\begin{tabular}{|c||c|c||c|c|c|c|c|c|c|c|c|c|}
\hline
 Model & \multicolumn{2}{c||}{Orthogonal $Q_{T}$} & 
 \multicolumn{10}{c|}{Block Bootstrap} \\
 & & & \multicolumn{2}{c|}{B=5} & \multicolumn{2}{c|}{B=10}
 &\multicolumn{2}{c|}{B=20} & \multicolumn{2}{c|}{B=30}
 &\multicolumn{2}{c|}{B=40}\\
\hline \hline           
       & 5\% & 10\% & 5\% & 10\% & 5\% & 10\% & 5\% & 10\% & 5\% & 10\%& 5\% & 10\%\\
\hline 
$X_{0.6,t}^{G}$ &   5.24 & 10.28 & 0 & 0 & 0.08 & 0.52 & 1.54 & 3.92 & 2.48 &
6 & 3.32 & 7.62 \\
$X_{0.6,t}^{\chi}$ & 5.2 & 9.78 & 0 & 0 & 0.02 & 0.16 & 0.52 & 1.84 & 0.04 &
0.14 & 2.12 & 6.68\\
$X_{0.9,t}^{\chi}$ & 2.24 & 4.84 &  0 & 0 & 0 & 0 & 0 & 0&0.04 & 0.62 & 0.04 & 0.62 \\
\hline 
\end{tabular}}
\caption{Goodness of fit test, under the null, T=500 over 5000 replications\label{table:7}}
\end{table}

\subsubsection*{Models under the alternative hypothesis}

To access the empirical power of the test we use realisations from the
same models considered in the null, namely $X_{0.6,t}^{G},
X_{0.6,t}^{\chi}$ and $X_{0.9,t}^{\chi}$ (their corresponding spectral
density functions are given in the previous section). To each of these models we
fit the spectral density function $g(\omega;\phi,\sigma) = (2\pi)^{-1}\sigma^{2}|1-\phi\exp(i\omega)|^{-2}$
for different values of $\phi$ and $\sigma$ (though $\sigma$ will
always be correctly specified). We used the sample sizes $T=100$, $T=200$ and $T=500$.

The result are given in Tables  \ref{table:8}, \ref{table:9} and \ref{table:10}.
For all the sample sizes considered, the power of the Block bootstrap
test increases with the block length,
though we recall that the largest block size the
type I errors were highly inflated. Overall, the power of the orthogonal sample
test is comparable (and often larger) than the power of the 
block bootstrap tests with the larger blocks. 

\begin{table}[h!]
\centering
\scalebox{0.8}{
\begin{tabular}{|c|c||c|c||c|c|c|c|c|c|c|c|c|c|}
\hline
 Model & Null&\multicolumn{2}{c||}{Orthogonal $Q_{T}$} & 
 \multicolumn{10}{c|}{Block Bootstrap} \\
 & & & & \multicolumn{2}{c|}{B=5} & \multicolumn{2}{c|}{B=10}
 &\multicolumn{2}{c|}{B=20} & \multicolumn{2}{c|}{B=30}
 &\multicolumn{2}{c|}{B=40} \\
\hline \hline           
   &    & 5\% & 10\% & 5\% & 10\% & 5\% & 10\% & 5\% & 10\% & 5\% & 10\% & 5\% & 10\%\\
\hline 

$X_{0.6,t}^{G}$ &$\sigma=1,\phi=0.3$& 60.4 & 71.24 & 1.98 & 11.32 & 35.42 & 55.12 & 53.24 & 70.78 &
61.2 & 76.5 & 68.62 & 80.34 \\

$X_{0.6,t}^{\chi}$ &$\sigma=\sqrt{2},\phi=0.3$ & 64.32 & 76.12 & 1.3 & 8.68 & 33.2 & 56.12 & 54.24 & 74.58 & 
62.4 & 80.42 & 71.06 & 84.14 \\

$X_{0.6,t}^{G}$ &$\sigma=1,\phi=0.45$ & 15.1 & 22.28 & 0 & 0 & 1.94 & 7.24 & 10.94 & 22.06 & 15.9& 28.62& 23.14& 36.36\\

$X_{0.6,t}^{\chi}$ &$\sigma=\sqrt{2},\phi=0.45$ & 13.98 & 22.38 & 0 & 0 & 0.82 & 3.96 & 6.42 & 16.78 & 10.72
& 23.6 &17.9 &32.74\\

$X_{0.9,t}^{\chi}$ &$\sigma=\sqrt{2},\phi=0.7$ & 41.14 & 51.82 & 0 & 0 & 3.2 & 10.28 & 17.54 & 36 & 25.54 &
45.58 & 36.24 & 56.44 \\ 
\hline
\end{tabular}}
\caption{Goodness of fit test, under the alternative, T=100 over 5000 replications\label{table:8}}
\end{table}

 \begin{table}[h!]
\centering
\scalebox{0.8}{
\begin{tabular}{|c|c||c|c||c|c|c|c|c|c|c|c|c|c|}
\hline
 Model & Null &\multicolumn{2}{c||}{Orthogonal $Q_{T}$} & 
 \multicolumn{10}{c|}{Block Bootstrap} \\
 & & & & \multicolumn{2}{c|}{B=5} & \multicolumn{2}{c|}{B=10}
 &\multicolumn{2}{c|}{B=20} & 
 \multicolumn{2}{c|}{B=30} &\multicolumn{2}{c|}{B=40} \\
\hline \hline           
    &   & 5\% & 10\% & 5\% & 10\% & 5\% & 10\% & 5\% & 10\% & 5\% &
       10\% & 5\% & 10\%\\
\hline
$X_{0.6,t}^{G}$ &$\sigma=1,\phi=0.3$ & 92.88 & 96.5 & 39.62 & 63.78 & 85.76 & 93.20 & 91.22 & 95.72
 & 93.36 & 97.04 & 93.48 & 97.40  \\
$X_{0.6,t}^{\chi}$ & $\sigma=\sqrt{2},\phi=0.3$ &96.04 & 98.5 & 33.5 & 62.1 & 88.86 & 96.02 & 93.9 & 98.4 &
95.14 & 98.80 & 95.66 & 98.92 \\
$X_{0.6,t}^{G}$ & $\sigma=1,\phi=0.45$& 38.38 & 49.5 & 0 & 0.1 & 8.88 & 20.38 & 27.16 & 42.2 & 33.52
& 48.28 & 39.08 & 53.82 \\
$X_{0.6,t}^{\chi}$& $\sigma=\sqrt{2},\phi=0.3$ & 40.46 & 53.08 & 0 & 0 & 6.04 & 15.86 & 21.28 & 38.98 &
27.34& 46.52 & 34.24 & 53.28\\
$X_{0.9,t}^{\chi}$ &$\sigma=\sqrt{2},\phi=0.7$ & 91.44 & 94.78 & 0 & 0 & 29.9 & 49.5 & 70.4 & 84.94 & 80.98 &
91 & 85.78 & 93.9 \\
\hline 
\end{tabular}}
\caption{Goodness of fit test, under the alternative, T=200 over 5000 replications\label{table:9}}
\end{table}

\begin{table}[h!]
\centering
\scalebox{0.8}{
\begin{tabular}{|c|c||c|c||c|c|c|c|c|c|c|c|c|c|}
\hline
 Model & Null &\multicolumn{2}{c||}{Orthogonal $Q_{T}$} & 
 \multicolumn{10}{c|}{Block Bootstrap} \\
& & & & \multicolumn{2}{c|}{B=5} & \multicolumn{2}{c|}{B=10}
 &\multicolumn{2}{c|}{B=20} 
& \multicolumn{2}{c|}{B=30} &\multicolumn{2}{c|}{B=40} \\
\hline \hline           
     &  & 5\% & 10\% & 5\% & 10\% & 5\% & 10\% & 5\% & 10\% & 5\% &
       10\% & 5\% & 10\%\\
\hline
 $X_{0.6,t}^{G}$ &$\sigma=1,\phi=0.3$& 99.98 & 100 & 98.88 & 99.72 & 100 & 100 & 100 & 100 & 100 &
100 & 100 & 100\\

$X_{0.6,t}^{\chi}$ &$\sigma=\sqrt{2},\phi=0.3$ & 99.98 & 100 & 99.2 & 99.92 & 100 & 100 & 100 & 100 & 99.92 &
100 &  99.92 & 100\\

$X_{0.6,t}^{G}$ &$\sigma=1,\phi=0.45$ & 86.90 & 92.52 & 0.18 & 1.54 & 48.88 & 65.62 & 77.36 & 86.7 & 81
& 89.93 &  82.58& 90.72\\

$X_{0.6,t}^{\chi}$ &$\sigma=\sqrt{2},\phi=0.45$ & 89.92 & 94.22 & 0.06 & 0.56 & 42.16 & 61.88 & 75.60 & 87.44 &
81.94 & 90.8 & 84.24 & 92.26 \\

$X_{0.9,t}^{G}$ &$\sigma=\sqrt{2},\phi=0.7$ & 99.96 & 99.98 & 0 & 0 & 94.46 & 98.12 & 99.86 & 99.96 & 99.9
& 99.98 & 99.92 & 99.98 \\
\hline 
\end{tabular}}
\caption{Goodness of fit test, under the alternative, T=500 over 5000 replications\label{table:10}}
\end{table}

%% file: 3_selection.tex
\section{Selection of $M$}\label{sec:cross}

In many respects, by using the $t_{2M}$-distribution to approximate
the distribution of $T_{M}$ in (\ref{eq:TM}) we are able to adjust to the choice of $M$.
However, one may view the choice of
$M$ as a little adhoc. 
Therefore, the purpose of 
this section is to propose an average square criterion for selecting
$M$.  Our proposed method is based on the results derived in Section \ref{sec:quad}. Using
Theorem \ref{lemma:2} we note that $\{\sqrt{T}A_{T}(\phi;r);1\leq r <T/2\}$ is
an 
almost uncorrelated,  zero mean sequence with variance 
$\var[\sqrt{T}A_{T}(\phi;r)] = V(\omega_{r})+O(T^{-1})$ (where $V(\omega_{r})$ is
defined in (\ref{eq:Vr})). These observations together with (\ref{eq:covV2}) imply that
\begin{eqnarray}
\left(\frac{|\sqrt{T}A_{T}(\phi;r)|^{2}}{V(\omega_{r})} - 1\right)
\qquad 1\leq r <T/2, \label{eq:AT}
\end{eqnarray}
is an almost uncorrelated sequence with mean zero and variance one. We
use this sequence as the building blocks of the average
criterion. In order to select $M$ we 
extend the estimator defined in (\ref{eq:estV}) to all frequencies
$V(\omega_{r})$. More precisely,
we use $\widehat{V}(\omega_{r})$ as an
estimator of $V(\omega_{r})$ where  
\begin{eqnarray*}
\widehat{V}_{M}(\omega_{r}) = \frac{T}{M}\sum_{s=1+r}^{M+r}|A_{T}(\phi;s)|^{2},
\end{eqnarray*}
noting that for $r=0$ we have the estimator defined in
(\ref{eq:estV}). Furthermore, by construction
$\widehat{V}_{M}(\omega_{r})$ is asymptotically uncorrelated to
$|\sqrt{T}A_{T}(\phi;r)|^{2}$ (see Theorem \ref{lemma:2}). The
suggested scheme is based on choosing the $M$ which minimises the
mean squared error of $\{\widehat{V}_{M}(\omega_{r});r=1,\ldots,T/p\}$.
This is analogous to bandwidth
selection in nonparametric regression. The choice of $p$ determines
over which frequencies the selection should be done (i) $p=2$ corresponds
to global bandwidth selection (ii) whereas large $p$ means focusing on 
estimation close to $r=0$. If there is not too much variability in the function
$V(\omega_{r})$ then the selection of $M$ should not be sensitive
to the choice of $p$. We define
 the average squared error 
\begin{eqnarray*}
\mathcal{C}_{\phi}(M) = \frac{p}{T}\sum_{r=1}^{T/p}\left(\frac{\left|\sqrt{T}A_{T}(\phi;r)\right|^{2}}{
\widehat{V}_{M}(\omega_{r})}-1 \right)^{2}.
\end{eqnarray*}
To select $M$ we use 
$\widehat{M} = \arg\min_{M\in \mathcal{S}}\mathcal{C}_{\phi}(M)$, where $\mathcal{S}$
is the set over which we do the selection. 

To understand what $\mathcal{C}_{\phi}(M)$ is estimating and why it
may be a suitable criterion, we make a Taylor
expansion of $\widehat{V}_{M}(\omega_{r})^{-1}$ about
$\Ex[\widehat{V}_{M}(\omega_{r})]^{-1}$ to give 
\begin{eqnarray*}
&&\left( \frac{|\sqrt{T}A_{T}(\phi;r)|^{2}}{\widehat{V}_{M}(\omega_{r})}
 -1 \right) \\
&\approx&
\frac{|\sqrt{T}A_{T}(\phi;r)|^{2} -
  \Ex[|\sqrt{T}A_{T}(\phi;r)|^{2}]}{\Ex[\widehat{V}_{M}(\omega_{r})]} 
- \frac{
  \Ex[|\sqrt{T}A_{T}(\phi;r)|^{2}]}{\Ex[\widehat{V}_{M}(\omega_{r})]^{2}}\left(\widehat{V}_{M}(\omega_{r})-\Ex[\widehat{V}_{M}(\omega_{r})]\right)
\\
&& +\left(\frac{\Ex[|\sqrt{T}A_{T}(\phi;r)|^{2}]}{\Ex[\widehat{V}_{M}(\omega_{r})]}-1\right).
\end{eqnarray*}
Taking the expectation squared of the above and using that
$|\sqrt{T}A_{T}(\phi;r)|$ and $\widehat{V}_{M}(\omega_{r})$ are
asymptotically uncorrelated we have 
\begin{eqnarray*}
&&\Ex\left( \frac{|\sqrt{T}A_{T}(\phi;r)|^{2}}{\widehat{V}_{M}(\omega_{r})}
 -1 \right)^{2} \\ 
&\approx&
\frac{\var[\sqrt{T}A_{T}(\phi;r)|^{2}]}{\Ex[\widehat{V}_{M}(\omega_{r})]^{2}} 
+ \left(\frac{
  \Ex[|\sqrt{T}A_{T}(\phi;r)|^{2}]^{}}{\Ex[\widehat{V}_{M}(\omega_{r})]^{2}}\right)^{2}
\underbrace{\var\left(\widehat{V}_{M}(\omega_{r})\right)}_{\textrm{variance}}
 +\underbrace{\left(\frac{\Ex[|\sqrt{T}A_{T}(\phi;r)|^{2}]}{\Ex[\widehat{V}_{M}(\omega_{r})]}-1\right)^{2}}_{\textrm{bias}}.
\end{eqnarray*}
Therefore $\mathcal{C}_{\phi}(M)$ is estimating 
\begin{eqnarray*}
\sum_{r=1}^{T/p}\left\{
\frac{\var[\sqrt{T}A_{T}(\phi;r)|^{2}]}{\Ex[\widehat{V}_{M}(\omega_{r})]^{2}} 
+ \left(\frac{
  \Ex[|\sqrt{T}A_{T}(\phi;r)|^{2}]^{}}{\Ex[\widehat{V}_{M}(\omega_{r})]^{2}}\right)^{2}
\var\left(\widehat{V}_{M}(\omega_{r})\right)+
 \left(\frac{\Ex[|\sqrt{T}A_{T}(\phi;r)|^{2}]}{\Ex[\widehat{V}_{M}(\omega_{r})]}-1\right)^{2}\right\}.
\end{eqnarray*}
Thus $\mathcal{C}_{\phi}(M)$ is balancing bias and variance. 
If $M$ is small, the bias is small but the variance is
large, conversely if $M$ large the bias is large but the variance
is small. It seems reasonable to choose the $M$ which balances these
two terms.  To illustrate how the
criterion behaves, in Figure \ref{fig:1} we
plot $\mathcal{C}_{\phi}(M)$ over $M$, for $\phi(\omega)   =
e^{i\omega}$ (which corresponds to the statistic which estimates the
autocovariance at lag one). As expected $\mathcal{C}_{\phi}(M)$ is
large when $M$ is both small and large. In this example, 
selecting $M$ anywhere between 7 and 13
seems to be reasonable.

%% file: appendix2.tex
\section*{Supplementary material}

To prove the results we make heavy use of the following well known
result. Suppose that the time series $\{X_{t}\}$ satisfies Assumption \ref{assum:p} with $p=s$.
Then we have
\begin{eqnarray}
\label{eq:bri}
\cum\left[J_{T}(\omega_{k_{1}}),\ldots,J_{T}(\omega_{k_{s}})\right] = 
\frac{1}{T^{s/2-1}}I_{k_{1}+\ldots+k_{s}\in
  T\mathbb{Z}}f_{s}(\omega_{k_{1}},\ldots,\omega_{k_{s-1}}) +
O(T^{-s/2}), 
\end{eqnarray}
see \cite{b:bri-81}, Theorem 4.3.2, for the details. 

\section{Proofs for Section \ref{sec:quad}}

{\bf PROOF of Lemma \ref{lemma:1}} This immediately follows from
(\ref{eq:bri}) and the Lipschitz continuity of $f$ and $\phi$ which allows the
sum to be replaced by an integral. \hfill $\Box$

\vspace{2mm}

\noindent {\bf PROOF of Theorem \ref{lemma:2}(i)} 
We first derive expressions for
$\cov[A_{T}(\phi;r_{1}),A_{T}(\phi;r_{2})] $ and
$\cov[A_{T}(\phi;r_{1}),\overline{A_{T}(\phi;r_{2})}]$ from which we
can deduce the covariance of $\Re A_{T}(\phi;r_{})$ and $\Im A_{T}(\phi;r_{})$.
 
To simplify notation we denote $J_{k} = J_{T}(\omega_{k})$. By using 
indecomposable partitions  and (\ref{eq:bri}) we have
\begin{eqnarray*}
&&T\cov[A_{T}(\phi;r_{1}),A_{T}(\phi;r_{2})] \\
&=&
\frac{1}{T^{}}\sum_{k_{1},k_{2}=1}^{T}
\phi(\omega_{k_{1}})\overline{\phi(\omega_{k_{2}})}
\cov[J_{k_{1}}\overline{J_{k_{1}+r_{1}}},J_{k_{1}}\overline{J_{k_{1}+r_{1}}}]
\\
 &=& \frac{1}{T^{}}\sum_{k_{1},k_{2}=1}^{T}
\phi(\omega_{k_{1}})\overline{\phi(\omega_{k_{2}})}\bigg(\cov[J_{k_{1}},J_{k_{2}}]\cov[\overline{J_{k_{1}+r_{1}}},\overline{J_{k_{2}+r_{2}}}]
  \\
&&+
\cov[J_{k_{1}},\overline{J_{k_{2}+r_{2}}}]\cov[\overline{J_{k_{1}+r_{1}}},J_{k_{2}}]
+
\cum[J_{k_{1}},\overline{J_{k_{1}+r_{1}}},\overline{J_{k_{2}}},J_{k_{2}+r_{2}}]
\bigg)\\
&=&\frac{1}{T^{}}\sum_{k=1}^{T}
\phi(\omega_{k_{}})\overline{\phi(\omega_{k_{}})}f(\omega_{k})f(\omega_{k+r_{1}})\delta_{r_{1}=r_{2}}
  +\frac{1}{T^{}}\sum_{k=1}^{T}
\phi(\omega_{k_{}})\overline{\phi(-\omega_{k_{}+r})}f(\omega_{k})f(\omega_{k+r_{1}})\delta_{r_{1}=r_{2}}\\
&&+
\frac{1}{T^{2}}\sum_{k_{1},k_{2}=1}^{T}\phi(\omega_{k_{1}})\overline{\phi(\omega_{k_{2}})}f_{4}(\omega_{k_{1}},-\omega_{k_{1}+r},-\omega_{k_{2}})\delta_{r_{1},r_{2}}
\bigg) + O(T^{-1}).
\end{eqnarray*}
Thus we see that if $r_{1}\neq r_{2}$, then
$|T\cov[A_{T}(\phi;r_{1}),A_{T}(\phi;r_{2})] | = O(T^{-1})$. On the
other hand if $r_{1} = r_{2}$ we replace the sum above with an
integral to give $T\var[A_{T}(\phi;r_{1})] = V(\omega_{r}) + O(T^{-1})$. 
We apply the same arguments to
$T\cov[A_{T}(\phi;r_{1}),\overline{A_{T}(\phi;r_{2})}]$ to give 
\begin{eqnarray*}
&&T\cov[A_{T}(\phi;r_{1}),\overline{A_{T}(\phi;r_{2})}] \\
&=&
\frac{1}{T^{}}\sum_{k_{1},k_{2}=1}^{T}
\phi(\omega_{k_{1}})\phi(\omega_{k_{2}})
\cov[J_{k_{1}}\overline{J_{k_{1}+r_{1}}},\overline{J_{k_{1}}}\overline{J_{k_{1}+r_{1}}}]
\\
 &=& \frac{1}{T^{}}\sum_{k_{1},k_{2}=1}^{T}
\phi(\omega_{k_{1}})\phi(\omega_{k_{2}})\bigg(\cov[J_{k_{1}},\overline{J_{k_{2}}}]\cov[\overline{J_{k_{1}+r_{1}}},J_{k_{2}+r_{2}}]
  \\
&&+
\cov[J_{k_{1}},\overline{J_{k_{2}+r_{2}}}]\cov[\overline{J_{k_{1}+r_{1}}},J_{k_{2}+r_{2}}]
+
\cum[J_{k_{1}},\overline{J_{k_{1}+r_{1}}},\overline{J_{k_{2}}},J_{k_{2}+r_{2}}]
\bigg) \\
&=&\frac{1}{T^{}}\sum_{k=1}^{T}
\phi(\omega_{k_{}})\phi(-\omega_{k_{}})f(\omega_{k})f(\omega_{k+r_{1}})\delta_{r_{1}=-r_{2}\textrm{
    or }T-r_{2}}
  +\frac{1}{T^{}}\sum_{k=1}^{T}
\phi(\omega_{k_{}})\phi(-\omega_{k_{}+r})f(\omega_{k})f(\omega_{k+r_{1}})
\delta_{r_{1}=-r_{2}\textrm{ or }T-r_{2}}\\
&&+
\frac{1}{T^{2}}\sum_{k_{1},k_{2}=1}^{T}\phi(\omega_{k_{1}})\phi(\omega_{k_{2}})f_{4}(\omega_{k_{1}},-\omega_{k_{1}+r},-\omega_{k_{2}})
\delta_{r_{1}=-r_{2}\textrm{ or }T-r_{2}}
\bigg) + O(T^{-1}).
\end{eqnarray*}
Since $0<r_{1},r_{2}<T/2$, the above implies that
$T\cov[A_{T}(\phi;r_{1}),\overline{A_{T}(\phi;r_{2})}] = O(T^{-1})$. 

Finally, by using the identities $\Re A_{T}(\phi;r) =
\frac{1}{2}(A_{T}(\phi;r_{1}) + \overline{A_{T}(\phi;r_{1})})$ and $\Im A_{T}(\phi;r) =
\frac{-i}{2}(A_{T}(\phi;r_{1}) - \overline{A_{T}(\phi;r_{1})})$ and
the above the result immediately follows. \hfill $\Box$

\vspace{2mm}
In order to prove  Theorem \ref{lemma:2}(ii) we require the following  
result. 
\begin{lemma}\label{lemma:cum}
Suppose that the time series $\{X_{t}\}$ satisfies Assumption \ref{assum:p} with $p=2n$.
Then we have 
\begin{eqnarray*}
\cum[\sqrt{T}A_{T}(\phi;r_{1}),\ldots,\sqrt{T}A_{T}(\phi;r_{n})] = O\left(\frac{1}{T^{n/2-1}}\right).
\end{eqnarray*}
\end{lemma}
{\bf PROOF} The proof immediately follows from using indecomposible
partitions and (\ref{eq:bri}). \hfill $\Box$

\vspace{2mm}
We make use of Lemma \ref{lemma:cum} below. 

\vspace{2mm}

\noindent  {\bf PROOF of Theorem \ref{lemma:2}(ii)} 
To prove (\ref{eq:covV2}) we note that 
\begin{eqnarray*}
&&\cov[|\sqrt{T}A_{T}(\phi;r_{1})|^{2},|\sqrt{T}A_{T}(\phi;r_{2})|^{2}] = \left|\cov[\sqrt{T}A_{T}(\phi;r_{1}),\sqrt{T}A_{T}(\phi;r_{2})]\right|^{2} \\
&& + \left|\cov[\sqrt{T}A_{T}(\phi;r_{1}),\sqrt{T}\overline{A_{T}(\phi;r_{2})}]
\right|^{2} + T^{2}\cum_{4}\left(A_{T}(\phi;r_{1}),\overline{A_{T}(\phi;r_{1})},A_{T}(\phi;r_{2}),\overline{A_{T}(\phi;r_{2})}\right).
\end{eqnarray*}
Thus we see that (\ref{eq:covV2}) follows immediately from the above, Theorem \ref{lemma:2} and Lemma \ref{lemma:4}.
\hfill $\Box$

\vspace{3mm}

\noindent {\bf PROOF of Lemma \ref{lemma:4}}
By using Corollary \ref{lemma:3} we can show that
\begin{eqnarray}
\label{eq:bias}
|\Ex[\widehat{V}_{M}(0)]-V(0)|=O(M/T).
\end{eqnarray}
To prove (\ref{eq:mse}) we use the classical variance bias
decomposition
\begin{eqnarray*}
\Ex\left(\widehat{V}_{M}(0) - V(0) \right)^{2} = \var[\widehat{V}_{M}(0)] + \left[\Ex[\widehat{V}_{M}(0)]-V(0)\right]^{2}.
\end{eqnarray*}
To bound $\var[\widehat{V}_{M}(0)]$ we note that
\begin{eqnarray*}
\var[\widehat{V}_{M}(0)] &=&
\frac{1}{M^{2}}\sum_{r_{1},r_{2}=1}^{M}\bigg\{\left|\cov[\sqrt{T}A_{T}(\phi;r_{1}),\sqrt{T}A_{T}(\phi;r_{2})]
\right|^{2} + \left|\cov[\sqrt{T}A_{T}(\phi;r_{1}),\sqrt{T}\overline{A_{T}(\phi;r_{2})}]
\right|^{2} \\
&& +
\cum\left(\sqrt{T}A_{T}(\phi;r_{1}),\sqrt{T}\overline{A_{T}(\phi;r_{1})},\sqrt{T}\overline{A_{T}(\phi;r_{2})},
\sqrt{T}A_{T}(\phi;r_{})\right)\bigg\} \\
 &=& O(M^{-1}+T^{-1}),
\end{eqnarray*}
where the last line follows immediately from (\ref{eq:covV2}).
Altogether by using the above and (\ref{eq:bias}), we obtain
desired result. \hfill $\Box$

\vspace{3mm}

\noindent {\bf PROOF of Lemma \ref{lemma:composite}} By using a Taylor expansion
we have 
\begin{eqnarray*}
A_{T}(\phi_{\widehat{\theta}};r) = A_{T}(\phi_{\theta};r) +
\left(\widehat{\theta}_{T}-\theta\right)A_{T}\left(\nabla_{\theta}\phi_{\theta};r\right)
 + \left(\widehat{\theta}_{T}-\theta\right)^{2}\frac{1}{T}\sum_{k=1}^{T}\nabla_{\theta}^{2}
 \phi(\omega_{k};\theta)\rfloor_{\theta = \bar{\theta}_{k}}J_{T}(\omega_{k})\overline{J_{T}(\omega_{k+r})},
\end{eqnarray*}
where $\bar{\theta}_{k}$ lies between $\theta$ and
$\widehat{\theta}$. Thus 
\begin{eqnarray*}
\left|A_{T}(\phi_{\widehat{\theta}};r)  - A_{T}(\phi_{\theta};r) -
\left(\widehat{\theta}_{T}-\theta\right)A_{T}\left(\nabla_{\theta}\phi_{\theta};r\right)\right|
  =O_{p}(T^{-2}).
\end{eqnarray*}
Therefore 
\begin{eqnarray}
\widehat{V}_{\widehat{\theta},M}(0) &=&
\frac{T}{M}\sum_{r=1}^{M}|A_{T}(\phi_{\theta};r)|^{2} +
\underbrace{|\widehat{\theta}_{T}-\theta|^{2}}_{O_{p}(T^{-1})}
\underbrace{\frac{T}{M}\sum_{r=1}^{M}|A_{T}(\nabla_{\theta}\phi_{\theta};r)|^{2}}_{=O_{p}(1),
\textrm{ by Lemma \ref{lemma:4}}} + O_{p}(T^{-1}) \nonumber\\
 &=& \frac{T}{M}\sum_{r=1}^{M}|A_{T}(\phi_{\theta};r)|^{2} +
 O_{p}(T^{-1}). \label{eq:6}
\end{eqnarray}
By using Lemma \ref{lemma:4} we have 
\begin{eqnarray*}
\Ex\left\{
\frac{T}{M}\sum_{r=1}^{M}|A_{T}(\phi_{\theta};r)|^{2} -
V_{\theta}(0)\right\}^{2} = O(M^{-1}+M/T), 
\end{eqnarray*}
thus by using this and (\ref{eq:6})
we obtain the result. 
\hfill $\Box$

\section{Proofs for Section \ref{sec:port}}

To prove Theorem \ref{theorem:orthogonalQ} we use the following definition 
\begin{eqnarray*}
V_{j_{1},j_{2}}(\omega_{r})  &=&
\frac{1}{2\pi}\int_{0}^{2\pi}f(\omega)f(\omega + \omega_{r})\left(
\phi_{j_{1}}(\omega)\overline{\phi_{j_{2}}(\omega)}
  +\phi_{j_{1}}(\omega)\overline{\phi_{j_{2}}(-\omega - \omega_{r})}
\right)
d\omega + \nonumber\\
&&\frac{1}{2\pi}\int_{0}^{2\pi}\int_{0}^{2\pi}\phi_{j_{1}}(\omega_{1})\overline{\phi_{j_{2}}(\omega_{2})}
f_{4}(\omega_{1},-\omega_{1}-\omega_{r},-\omega_{2})d\omega_{1}d\omega_{2}+ O(T^{-1}).
\end{eqnarray*}
The following lemma facilitates the proof of Theorem \ref{theorem:orthogonalQ}.
\begin{lemma}
Suppose that the time series $\{X_{t}\}$ satisfies Assumption \ref{assum:p} with $p=16$.
Then we have 
\begin{eqnarray}
T\cov\left[A_{T}(\phi_{j_{1}};0),A_{T}(\phi_{j_{2}};0)\right] &=&
V_{j_{1},j_{2}} + O(T^{-1})\nonumber\\
T\cov\left[A_{T}(\phi_{j_{1}};0),\overline{A_{T}(\phi_{j_{2}};0)}\right] &=&
V_{j_{1},j_{2}} + O(T^{-1}) \label{eq:F1}.
\end{eqnarray}
For  all $0<r_{1},r_{2}<T/2$ we have 
\begin{eqnarray}
T\cov\left[A_{T}(\phi_{j_{1}};r_{1}),A_{T}(\phi_{j_{2}};r_{2})\right] &=&
\left\{
\begin{array}{cc}
V_{j_{1},j_{2}}(\omega_{r}) + O(T^{-1}) & r_{1}  = r_{2}\\
 O(T^{-1}) & r_{1} \neq r_{2}\\
\end{array}
\right. \nonumber\\
T\cov\left[A_{T}(\phi_{j_{1}};r_{1}),\overline{A_{T}(\phi_{j_{2}};r_{2})}\right] &=&
 O(T^{-1}) \label{eq:F2}.
\end{eqnarray}
For all $0\leq r_{1},r_{2},r_{3},r_{4}<T/2$ we have 
\begin{eqnarray}
T^{2}\cum\left[A_{T}(\phi_{j_{1}};r_{1}),A_{T}(\phi_{j_{2}};r_{2}),A_{T}(\phi_{j_{3}};r_{3}),A_{T}(\phi_{j_{4}};r_{4})
\right] = O(T^{-1}). \label{eq:cum4n}
\end{eqnarray}
For $0<r<T/2$ and $0<r_{1},r_{2}<T/2$
we have
\begin{eqnarray}
T^{2}\cov[A_{T}(\phi_{j_{1}};r_{1})^{2},A_{T}(\phi_{j_{2}};r_{2})^{2}]
&=& 
\left\{
\begin{array}{cc}
2V_{j_{1},j_{2}}(\omega_{r})^{2} + O(T^{-1}) & r_{1} = r_{2} (=r)\\
 O(T^{-1}) & r_{1}\neq r_{2}
\end{array}
\right. \label{eq:cum4-all1}\\
T^{2}\cov[A_{T}(\phi_{j_{1}};r_{1})^{2},A_{T}(\phi_{j_{2}};r_{2})\overline{A_{T}(\phi_{2};r_{2})}]
&=& O(T^{-1}) \label{eq:cum4-all2}\\
T^{2}\cov[|A_{T}(\phi_{j_{1}};r_{1})|^{2},|A_{T}(\phi_{j_{2}};r_{2})|^{2}]
&=& 
\left\{
\begin{array}{cc}
V_{j_{1},j_{2}}(\omega_{r})^{2} +O(T^{-1}) \\
O(T^{-1})\\
\end{array}
\right. \label{eq:cum4-all3} 
\end{eqnarray}
Finally 
\begin{eqnarray}
\left| V_{j_{1},j_{2}}(\omega_{r}) - V_{j_{1},j_{2}} \right| \leq
K|r|T^{-1}, \label{eq:Vj1j2}
\end{eqnarray}
where $K$ is a finite constant. 
\end{lemma}
{\bf PROOF}  The proof of (\ref{eq:F1}) and (\ref{eq:F2}) is identical
to the proof of Theorem \ref{lemma:2}, thus we omit the details. 
The proof of (\ref{eq:cum4n}) follows from Lemma
\ref{lemma:cum}, thus we omit the details. 
To prove (\ref{eq:cum4-all1}) we use indecomposable partitions to
decompose the term in the product of covariances and a fourth order
cumulant term. Specifically 
\begin{eqnarray*}
&&T^{2}\cov[A_{T}(\phi_{j_{1}};r_{})^{2},A_{T}(\phi_{j_{2}};r_{2})^{2}] \nonumber\\
&=&
2T^{2}|\cov[A_{T}(\phi_{j_{1}};r_{1}),A_{T}(\phi_{j_{1}};r_{2})]|^{2}
+ T^{2}\cum[A_{T}(\phi_{j_{1}};r_{1}),A_{T}(\phi_{j_{1}};r_{1}),\overline{A_{T}(\phi_{j_{1}};r_{2})},\overline{A_{T}(\phi_{j_{1}};r_{2})}].
\end{eqnarray*}
By using (\ref{eq:F2}) we obtain  (\ref{eq:cum4-all1}). A similar
proof applies to (\ref{eq:cum4-all2}) and (\ref{eq:cum4-all3}). 

Finally, to prove (\ref{eq:Vj1j2}) we simply use the Lipschitz
continuity of $f$, $f_{4}$ and $\phi_{j}$. Thus we have proved
the result. \hfill $\Box$

\vspace{2mm}

\noindent {\bf PROOF of Theorem \ref{theorem:orthogonalQ}} To prove (i), we
note that under 
both the null and alternative the following
expansion is valid
\begin{eqnarray}
\Ex[S_{T}] &=& T\sum_{j=1}^{L}\Ex\left|\left(A_{T}(\phi_{j_{}})
    -\Ex[A_{T}(\phi_{j})\right) +
  \Ex[A_{T}(\phi_{j})]\right|^{2}\nonumber\\
 &=& T\sum_{j=1}^{L}\Ex\left|\left(A_{T}(\phi_{j})
    -\Ex[A_{T}(\phi_{j})\right) +  \Ex[A_{T}(\phi_{j})]\right|^{2} \nonumber\\
 &=& T\sum_{j=1}^{L}\var[A_{T}(\phi_{j})] +
 \sum_{j=1}^{L}\left|\Ex[A_{T}(\phi_{j})] \right|^{2}.\label{eq:EQ}
\end{eqnarray}
Using that under the null $\Ex[A_{T}(\phi_{j})] =0$, and
substituting this into the above we have 
\begin{eqnarray*}
\Ex[S_{T}] &=& T\sum_{j=1}^{L}\var[\sqrt{T}A_{T}(\phi_{j})] =
\sum_{j=1}^{L}V_{j,j} + O(T^{-1}),
\end{eqnarray*}
where the last line follows from (\ref{eq:F2}). This gives (ia). To
prove (ib) we note that 
\begin{eqnarray*}
\Ex[S_{T,R}(r)] =
\frac{T}{2}\sum_{j=1}^{L}\var\left[A_{T}(\phi_{j};r) +
  \overline{A_{T}(\phi_{j};r)}\right] + \frac{T}{2}\sum_{j=1}^{L} \left|\Ex\left(A_{T}(\phi_{j};r) +
  \overline{A_{T}(\phi_{j};r)}\right)\right|^{2}.
\end{eqnarray*}
Under both the null and alternative
$\Ex[A_{T}(\phi_{j};r)]=O(T^{-1})$ 
for $0<r<T/2$. Thus 
\begin{eqnarray*}
\Ex[S_{T,R}(r)] &=&
\frac{T}{2}\sum_{j=1}^{L}\left(
2\var[\sqrt{T}A_{T}(\phi_{j};r)] +2\Re\cov[\sqrt{T}A_{T}(\phi_{j};r),\overline{\sqrt{T}A_{T}(\phi_{j};r)}]
\right) + O(T^{-1}) \\
&=& \sum_{j=1}^{L}V_{j,j} + O(T^{-1}),
\end{eqnarray*}
thus proving (ib).

To prove (iia) we note that 
since $A_{T}(\phi_{j})$ is real and under the null
$\Ex[A_{T}(\phi_{j})]$, then expanding $\var[S_{T}]$ gives
\begin{eqnarray*}
\var[S_{T}] 
&=&
T^{2}\sum_{j_{1},j_{2}=1}^{L}\cov\left(|\sqrt{T}A_{T}(\phi_{j_{1}})|^{2},|\sqrt{T}A_{T}(\phi_{j_{2}})|^{2}\right)
\\
 &=&
 T^{2}\sum_{j_{1},j_{2}=1}^{L}\bigg(2\cov[\sqrt{T}A_{T}(\phi_{j_{1}}),\sqrt{T}A_{T}(\phi_{j_{2}})]^{2}
 +\\
&& T^{2}\cum[A_{T}(\phi_{j_{1}}),A_{T}(\phi_{j_{1}}),A_{T}(\phi_{j_{1}}),A_{T}(\phi_{j_{2}}),A_{T}(\phi_{j_{1}})]\bigg)
\\ 
&=&  2\sum_{j_{1},j_{2}=1}^{L}V_{j_{1},j_{2}}^{2} + O(T^{-1}),
\end{eqnarray*}
where the last line follows from (\ref{eq:F2}) and
(\ref{eq:cum4n}). Thus proving (iia).

We now prove (iib), where we derive an expression for  
$\cov[|\Re A_{T}(\phi_{j_{1}};r)|^{2},|\Re
A_{T}(\phi_{j_{2}};r)|^{2}]$. To simplify notation let $A_{T}(r)=A_{T}(\phi_{j};r)$.
Using this notation we write $\Re
A_{T}(\phi_{j};r) = \frac{1}{2}(A_{T}(r) +
\overline{A_{T}(r)})$ and 
\begin{eqnarray*}
|\Re A_{T}(\phi_{j};r)|^{2} = \frac{1}{4}\left(A_{T}(r)^{2} +
  A_{T}(r)\overline{A_{T}(r)} + \overline{A_{T}(r)}A_{T}(r) + \overline{A_{T}(r)}^{2} \right).
\end{eqnarray*}
Thus 
\begin{eqnarray*}\cov[|\Re A_{T}(\phi_{j};r_{1})|^{2},|\Re
A_{T}(\phi_{j};r_{2})|^{2}]  
&=& \frac{1}{16}
\cov\big[A_{T}(r_{1})^{2} +
  A_{T}(r_{1})\overline{A_{T}(r_{1})} +
  \overline{A_{T}(r_{1})}A_{T}(r_{1}) + \overline{A_{T}(r_{1})}^{2},\nonumber\\
&& A_{T}(r_{2})^{2} +
  A_{T}(r_{2})\overline{A_{T}(r_{2})} + \overline{A_{T}(r_{2})}A_{T}(r_{2}) + \overline{A_{T}(r_{2})}^{2} 
\big].
\end{eqnarray*}
Thus by using (\ref{eq:cum4-all1})-(\ref{eq:cum4-all3}) we have 
\begin{eqnarray*}
\cov[|\sqrt{T}\Re A_{T}(\phi_{j};r_{1})|^{2},|\sqrt{T}\Re
A_{T}(\phi_{j};r_{2})|^{2}]  =
\left\{
\begin{array}{cc}
\frac{1}{2}\sum_{j_{1},j_{2}=1}^{L}V_{j_{1},j_{2}}(\omega_{r}) + O(T^{-1}) & r_{1}
= r_{1}\\
O(T^{-1}) & r_{1}\neq r_{2}
\end{array} 
\right. 
\end{eqnarray*}
Now we recall that $S_{T,R}(r) = 2T\sum_{j=1}^{L}|\Re
A_{T}(\phi_{j};r)|^{2} $, thus by using the above and (\ref{eq:Vj1j2}) we have 
\begin{eqnarray*}
\cov[S_{T,R}(r_{1}),S_{T,R}(r_{2})] = 
\left\{
\begin{array}{cc}
2\sum_{j_{1},j_{2}=1}^{L}V_{j_{1},j_{2}} + O(T^{-1} + |r|T^{-1}) &
r_{1} = r_{2} (=r) \\
O(T^{-1}) & r_{1}\neq r_{2}.
\end{array}
\right.
\end{eqnarray*}
The same arguments apply to $\cov[S_{T,R}(r_{1}),S_{T,R}(r_{2})]$ and
$\cov[S_{T,R}(r_{1}),S_{T,I}(r_{2})]$, which gives us (iib). 

To prove (iii) we use the same method used to prove (ii) together with Lemma \ref{lemma:cum}. 
\hfill $\Box$

\vspace{3mm}